\documentclass[useAMS,usenatbib]{mn2e}
\usepackage{graphicx} 
\usepackage{threeparttable}
\usepackage{amssymb}
\usepackage{longtable}
\usepackage{natbib}
\usepackage{amsmath} 
\usepackage{ulem} 

\title[Frequency and brightness of white dwarf discs]{The frequency and infrared brightness of circumstellar discs at white dwarfs}

\author[M. Rocchetto et al.]{M. Rocchetto$^{1}$\thanks{E-mail:
m.rocchetto@ucl.ac.uk}, J. Farihi$^{1}$\thanks{STFC Ernest Rutherford Fellow}, B. T. G{\"a}nsicke$^{2}$,  C. Bergfors$^{1}$\\
$^{1}$Department of Physics \& Astronomy, University College London, Gower Street, London, WC1E 6BT, UK\\
$^{1}$Department of Physics, University of Warwick, Coventry CV4 7AL, UK}

\begin{document}
\normalem
\date{Accepted 2015 February 10. Received 2015 January 15; in original form 2014 October 21}

\pagerange{\pageref{firstpage}--\pageref{lastpage}} \pubyear{2014}

\maketitle

\label{firstpage}

\begin{abstract}
White dwarfs whose atmospheres are polluted by terrestrial-like planetary debris have become a powerful and unique tool to study evolved planetary systems. This paper presents results for an unbiased \emph{Spitzer} IRAC search for circumstellar dust orbiting a homogeneous and well-defined sample of 134 single white dwarfs. The stars were selected without  regard to atmospheric metal content but were chosen to have 1) hydrogen rich atmospheres, 2) \mbox{$17\,000$\,K $ < T_\mathrm{eff} < 25\,000$\,K} and correspondingly young post main-sequence ages of 15--270\,Myr, and 3) sufficient far-ultraviolet brightness for a corresponding \emph{Hubble Space Telescope}  COS Snapshot.  Five white dwarfs were found to host an infrared bright dust disc, three previously known, and two reported here for the first time, yielding a nominal  $3.7^{+2.4}_{-1.0}$\% of white dwarfs in this post-main sequence age range with detectable circumstellar dust.  Remarkably, the complementary \emph{Hubble} observations indicate that a fraction of 27\% show metals in their photosphere that can only be explained with ongoing accretion from circumstellar material, indicating that nearly 90\% of discs escape detection in the infrared, likely due to  small emitting surface area. This paper also presents the distribution of disc fractional luminosity as a function of  cooling age for all known dusty white dwarfs, suggesting possible disc evolution scenarios and indicating an undetected population of circumstellar discs. 

\end{abstract}


\section{Introduction}

The field of exoplanetary research is one of the most rapidly expanding subjects in astrophysics. The large number of new discoveries made in the past 20 years have led to swift progress and a preliminary characterisation of the Galactic planet population. Current detection techniques enable estimates of exoplanet radius and mass, using mainly the transit and radial velocity methods, but density alone does not yield a unique solution for the bulk composition of the planet \citep*{2008ApJ...673.1160A,2007ApJ...665.1413V}. Transit spectroscopy can also provide some insights into the composition of exoplanet atmospheres \citep{2007ApJ...658L.115G,2007Natur.448..169T} but to date has proved difficult. 

In this context, white dwarfs offer a unique laboratory to study exoplanetary compositions. It is now clear that  planetary systems around Sun-like and intermediate-mass stars survive, at least in part, the post-main sequence phases of their hosts \citep{2010ApJ...722..725Z}.  Compelling evidence comes from metal polluted white dwarfs that commonly exhibit closely orbiting circumstellar dust (and sometimes gas) discs originating from the disruption of large asteroids or  planetesimals (\citealp*{2008ApJ...674..431F}; \citealp{2007ApJ...662..544V}; \citealp{2006Sci...314.1908G}; \citealp{2005ApJ...632L.119B}; \citealp{1987Natur.330..138Z}).  

Owing to high surface gravity and negligible radiative forces, heavy elements sink on relatively short timescales within the atmospheres of relatively cool ($T_\mathrm{eff} \lesssim 25\,000$\,K) white dwarfs if compared to the evolution timescales \citep{1986ApJS...61..197P,1979ApJ...231..826F}. Hence, the presence of metals in the atmospheres of cool degenerates must be a sign of recent external accretion \citep{2003ApJ...596..477Z}. The source of this accreting material was initially attributed to the interstellar medium (\citealp{1993ApJS...84...73D}; \citealp*{1993ApJS...87..345D}; \citealp{1992ApJS...82..505D}) or comets \citep*{1986ApJ...302..462A}, but both theories had trouble in explaining the high and ongoing accretion rates found at hydrogen dominated white dwarfs \citep{2003ApJ...596..477Z}. Today  accretion from circumstellar material, resulting from the disruption of large asteroids or minor planets, is, by far, the most compelling explanation for atmospheric metals seen at a large fraction of cool white dwarfs \citep{2013MNRAS.431.1686V,2003ApJ...584L..91J}.  

Metal enriched white dwarfs have become a powerful tool to indirectly analyse the composition of exoterrestrial planetary matter, as their photospheres in principle mirror the composition of the accreted material.  As an example of this technique, it was demonstrated that the relative abundances of 15 heavy elements in the atmosphere of GD\,362  reflect the composition of a large asteroid that was similar in composition to the bulk Earth-Moon system \citep{2007ApJ...671..872Z}.  Notably, ultraviolet and optical spectroscopy have shown that metal-contaminated degenerates are, in general, refractory-rich and volatile-poor (\citealp{2012MNRAS.424..333G}; \citealp{2010ApJ...709..950K}; \citealp{2008ApJ...672..540D}; \citealp{2007ApJ...663.1291D}; \citealp*{2002A&A...385..995W}), while infrared spectroscopy reveals that the circumstellar dust itself is silicate-rich and carbon-poor (\citealp*{2009AJ....137.3191J}; \citealp{2009ApJ...693..697R}; \citealp{2005ApJ...635L.161R}), and thus similar to materials found in the inner Solar System \citep{2003ApJ...591.1220L}. Furthermore,   a circumstellar disc that resulted from the destruction of  a rocky and water-rich extrasolar minor planet was identified around the white dwarf GD\,61 \citep*{2013Sci...342..218F}, demonstrating the existence of water in terrestrial zone planetesimals that could play an important role in delivering water to the surface of planets. 

Detailed modelling of the infrared excesses found at a fraction of metal polluted white dwarfs suggests that the circumstellar dust is arranged in the form of an optically thick but geometrically thin disc, with similar properties to the rings of Saturn  \citep{2006ApJ...646..275R,2003ApJ...584L..91J}. These rings of warm dust are situated within the Roche limit of their host star, as also confirmed by the emission and absorption profiles of gaseous debris discovered at several dusty white dwarfs \citep{2012ApJ...750...86B,2012ApJ...754...59D,2009ApJ...696.1402B,2006Sci...314.1908G}. There is substantial theoretical support for disc creation via tidal disruption of post-main sequence planetary systems, perturbed by unseen planets \citep{2014arXiv1409.2493V,2013MNRAS.431.1686V,2002ApJ...572..556D}. The transition from disruption to disc is still poorly understood but there are good models for the evolution of these metal dominated discs (\citealp{2014MNRAS.439.2442F}; \citealp*{2012ApJ...747..148D}; \citealp*{2012MNRAS.423..505M}; \citealp*{2011MNRAS.414..930B}).

After nearly a decade of dust disc discoveries at metal enriched degenerates, the statistical frequency of the phenomenon still suffers from significant observational biases. While the fraction of metal polluted white dwarfs that are currently accreting has been constrained by several unbiased surveys to about 20--30\%  (\citealp*{2014A&A...566A..34K}; \citealp{2010ApJ...722..725Z}; \citealp{2003ApJ...596..477Z}), surveys aiming to detect infrared bright dust discs at white dwarfs suffer from considerable biases. The first searches for infrared excesses targeted relatively cool ($T_\mathrm{eff} \lesssim 25\,000$\,K)  stars known to be metal-polluted (\citealp{2010ApJ...714.1386F}; \citealp*{2007ApJ...663.1285J}; \citealp*{2007AJ....134.1662D}), and this approach does not permit robust statistics of disc frequency over the entire white dwarf population. With little restriction of stellar effective temperature, wide field surveys such as SDSS, UKIDSS, and \emph{WISE} have found disc frequencies between 0.4 and 1.9\% \citep{2011MNRAS.417.1210G,2011MNRAS.416.2768S,2011ApJS..197...38D}. On the other hand, more sensitive \emph{Spitzer} observations of $K$-band bright white dwarfs, without regard to stellar temperature, resulted in a nominal disc frequency of 1.6\% \citep{2007ApJS..171..206M}. A higher frequency of 4.5\% was obtained by \cite{2012ApJ...760...26B}, by targeting stars in a restricted temperature range where dust detections were expected based on prior surveys. Moreover, the observed sample was fragmented over several instruments and hence the result is difficult to compare with other surveys. 

In order to determine definite statistics of the frequency of infrared bright discs at white dwarfs, this paper presents \emph{Spitzer} observations of a well defined sample of 134 young DA white dwarfs in the temperature range $17\,000\,\mathrm{K} < T_\mathrm{eff} < 25\,000$\,K.  Metal abundances for 85 of the stars were also determined in a complementary \emph{Hubble Space Telescope} COS Snapshot program \citep{2014A&A...566A..34K}, and permit the first unbiased statistics of the frequency of circumstellar discs at young white dwarfs. 

Section 2 describes the sample selection criteria and \emph{Spitzer} photometry, Section 3 discusses how the spectral energy distributions of the sample stars were obtained and describes those with detected discs. Sections 4 and 5 present the derived disc frequency in the context of complementary \emph{HST} ultraviolet observations and previous surveys, and Section 6 discusses the distribution of the  fractional disc luminosity for all known dusty white dwarfs. Section 7 presents notes on the individual objects, and Section 8 gives a short conclusion.

\section{Observations}

\subsection{Sample selection}

\begin{figure*}
\includegraphics{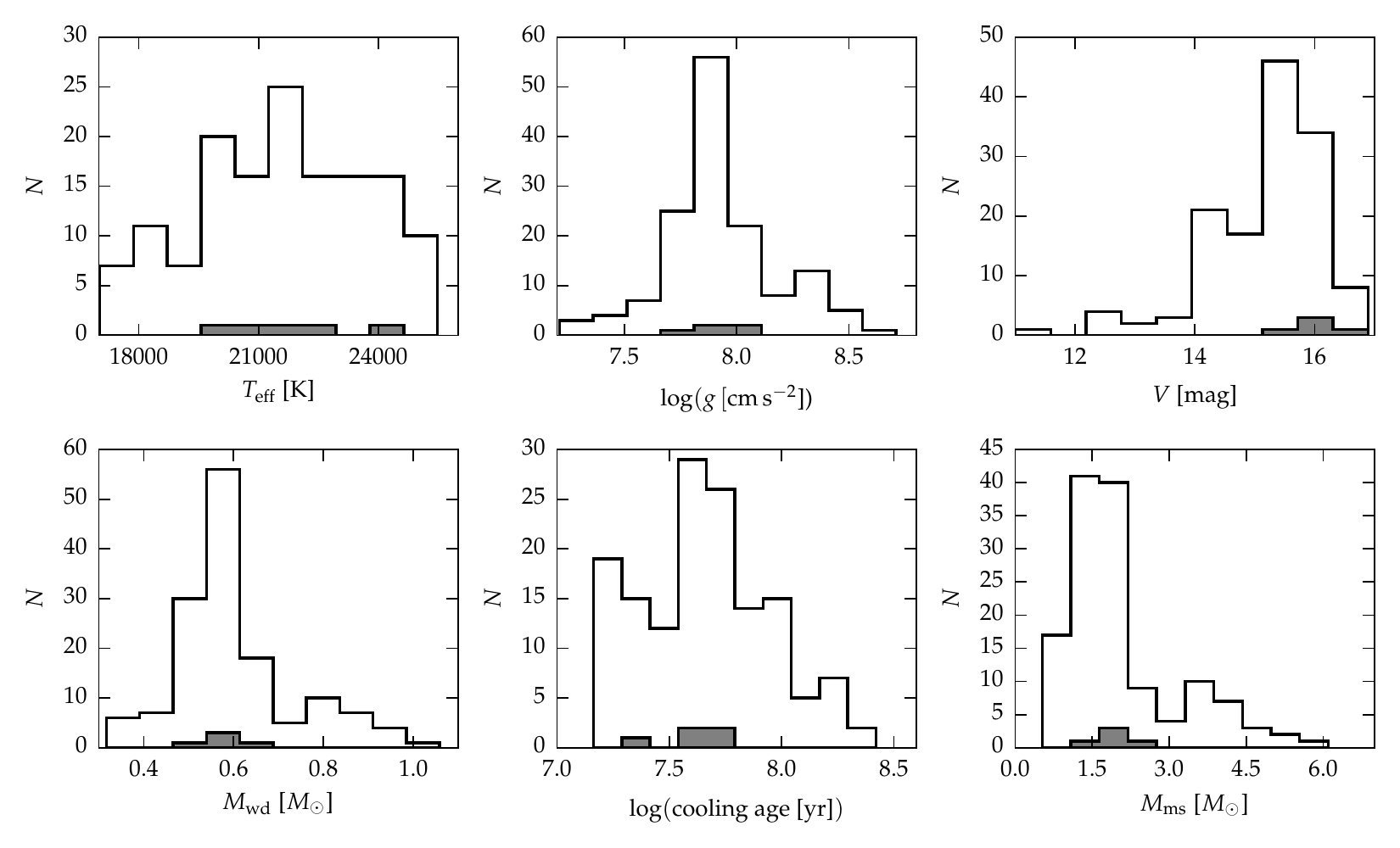}
\caption[foo]{Fundamental properties of the 134 DA stars observed in the \emph{Spitzer} survey. Effective temperature $T_\mathrm{eff}$ and surface gravity $\log g$ are taken from \protect{\cite{2005ApJS..156...47L}} or \protect{\cite{2009A&A...505..441K}}, and $V$-band magnitudes are taken from the AAVSO Photometric All-Sky Survey (APASS). Final white dwarf mass $M_\mathrm{wd}$ and cooling age are derived using evolutionary cooling sequences \citep*{2001PASP..113..409F}; the initial main-sequence progenitor mass $M_\mathrm{ms}$ is derived from $M_\mathrm{wd}$ and the initial-to-final mass relation \protect{\citep{2008ApJ...676..594K}}. The entire sample of white dwarfs is shown with the unfilled histograms, while the degenerates with infrared detected discs are shown in grey.}
\label{fig:star_properties}
\end{figure*}

The sample of hydrogen dominated white dwarfs was selected from the catalogues compiled by  \cite*{2005ApJS..156...47L} and \cite{2009A&A...505..441K}, who performed model atmosphere analyses based on optical spectroscopy, providing effective temperature, surface gravity, mass and cooling age. The only criteria for selection were 1)  $17\,000\,\mathrm{K} < T_\text{eff} < 25\,000$\,K  and corresponding young cooling ages of 15--270\,Myr, 2) predicted fluxes $F_\lambda(1300\,\text{\AA}) > 5 \times 10^{-14}\,\text{erg\,cm}^{-2}\,\text{s}^{-1}\text{\AA}^{-1}$  for the corresponding \emph{HST} COS Snapshot survey. As such, the selection was performed only for temperature and brightness, resulting in an unbiased sample of 134 DA young single white dwarfs. The distributions of the fundamental stellar parameters are shown in Figure \ref{fig:star_properties}.

\subsection{Spitzer observations}

A total of 100 sample stars were observed  between 2012 May and October in the 3.6 and 4.5\,$\mu$m bandpasses using the Infrared Array Camera \citep[IRAC;][]{2004ApJS..154...10F} on-board the  \emph{Spitzer Space Telescope} \citep{2004ApJS..154....1W} as part of Program 80149\footnote{Five stars were also observed as part of the same Program but are not included in this study. The white dwarf 1929+012 was observed as an ancillary target (see Section \ref{sec:1929}) while four white dwarfs (0933+025, 1049+103, 1335+369, 1433+538) initially considered to be single stars and included in the original sample were found to host unresolved M dwarf companions and were therefore excluded. Infrared fluxes for these four binaries are reported in the Appendix.}.  An exposure time of 30\,s was used for each individual frame, with 20 medium-size dithers in the cycling pattern, resulting in 600\,s total exposure time in each warm IRAC channel. The remaining 34 sample stars were previously observed during either the cold or warm mission, and their archival data were analysed.

The analysis of all  targets was performed using the \mbox{$0\farcs6$  pixel$^{-1}$} mosaics (Post-Basic Calibrated Data) processed by the IRAC calibration pipeline version S\,18.25.0 which produces a single, fully processed and calibrated image.  Aperture photometry was performed using the point source extraction package {\sc apex} within {\sc mopex} \citep{2006SPIE.6274E..10M} and an aperture radius of 4 pixels with a 24--40 pixels sky annulus. Fluxes were corrected for aperture size, but not for colour. For blended sources, point response function (PRF) fitting was performed on the Basic Calibrated Data frames, using the package {\sc apex multiframe} within {\sc mopex}. Fluxes obtained with PRF fitting were compared with the   {\sc daophot} PSF fitting routine within {\sc iraf} for a representative sample, and led to consistent results to within 4\%, and always within the relative uncertainties. The measured flux uncertainty was computed by {\sc apex} and include the source photon noise and the variance in sky background. A 5\% calibration uncertainty is conservatively added in quadrature  to all IRAC fluxes. The flux determinations and uncertainties for the science targets, together with the physical parameters of the stars, are reported in the Appendix.

\subsection{Additional near-infrared observations}

Independent $JHK_\mathrm{s}$ photometry was also obtained for five targets (Table \ref{tab:jhk_phot}). The stars were observed in 2011 October with LIRIS \citep{1998SPIE.3354..448M} at the William Herschel Telescope (WHT) and in 2011 August with SOFI \citep{1998Msngr..91....9M} at the New Technology Telescope (NTT), with exposure times of 30, 15, and 10\,s at $J$, $H$, and $K_s$ respectively.  Sufficient dithers were performed for a total exposure time of 270 s at each bandpass for each target. Three standard star fields from the ARNICA catalogue \citep{1998AJ....115.2594H} were observed each night for flux calibration. The data were reduced using standard aperture photometry with the {\sc iraf} task {\sc apphot}. The flux calibration for all nights was good to 5\% or better.

\begin{table}
  \caption{Independent $JHK_\mathrm{s}$ photometry. All errors are 5\%.}
  \begin{tabular}{@{}lllll@{}}
  \hline
WD & $J$   & $H$   & $K_\mathrm{s}$ & Instrument\\
   & (mag) & (mag) & (mag)          & /Telescope \\
\hline
1013+256 & $16.90$ & $17.03$ & $17.15$ & LIRIS/WHT\\
0843+516 & $16.56$ & $16.58$ & $16.49$ & LIRIS/WHT\\
0431+126 & $14.79$ & $14.83$ & $14.94$ & LIRIS/WHT\\
0421+162 & $14.81$ & $14.83$ & $14.89$ & LIRIS/WHT\\
1953--175 & $15.59$ & $15.68$ & $15.73$ & SOFI/NTT\\
\hline
\end{tabular}
\label{tab:jhk_phot}
\end{table}

\section{Data Analysis}

\label{sect:data-analysis}

The spectral energy distributions (SEDs) of the sample stars were constructed with additional short wavelength photometry from a variety of catalogues, including Sloan Digital Sky Survey \citep[SDSS;][]{2012ApJS..203...21A}, AAVSO Photometric All-Sky Survey \citep[APASS;][]{2009AAS...21440702H}, Two Micron All Sky Survey \citep[2MASS;][]{2006AJ....131.1163S}, UKIRT Infrared Deep Sky Survey \citep[UKIDSS;][]{2007MNRAS.379.1599L}, and Deep Near Infrared Survey of the Southern Sky \citep[DENIS;][]{1999A&A...349..236E}. Additional near-infrared fluxes for a few targets were also obtained from the literature \citep{2012ApJ...760...26B,2009MNRAS.398.2091F}.

The available optical and near-infrared fluxes were fitted with pure hydrogen white dwarf atmosphere models \citep{2010MmSAI..81..921K}, kindly provided by the author. The fit was computed by matching the optical and near-infrared best-quality photometric data points with a model spectrum with \mbox{$\log g = 8$} and effective temperature obtained  from \cite{2005ApJS..156...47L} or \cite{2009A&A...505..441K}, approximated to the closest available model.  The Levenberg-Marquardt minimisation algorithm was used to find the best scaling factor.  

\subsection{Stars with infrared excesses}

\label{sec:infrared_excesses}
 
\begin{figure*}
\vspace*{1cm}
\includegraphics{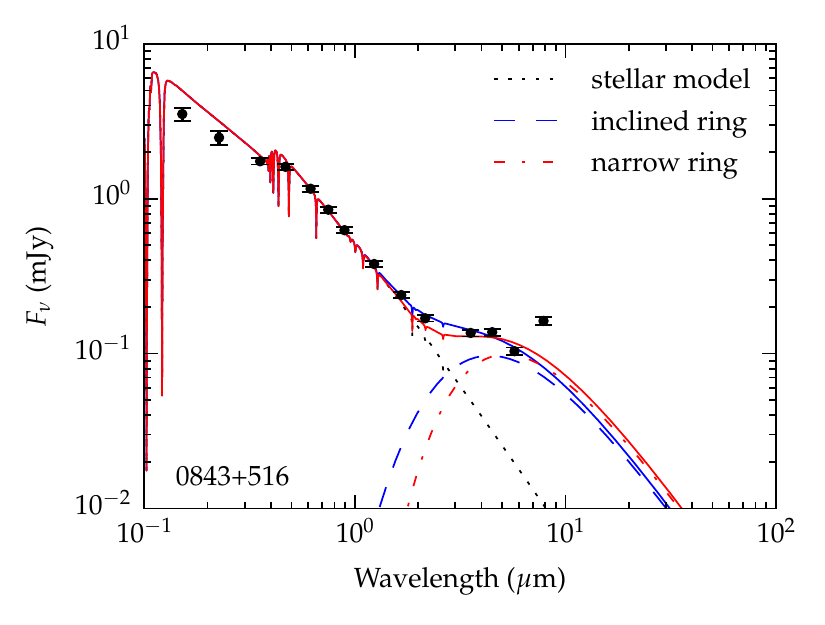}
\includegraphics{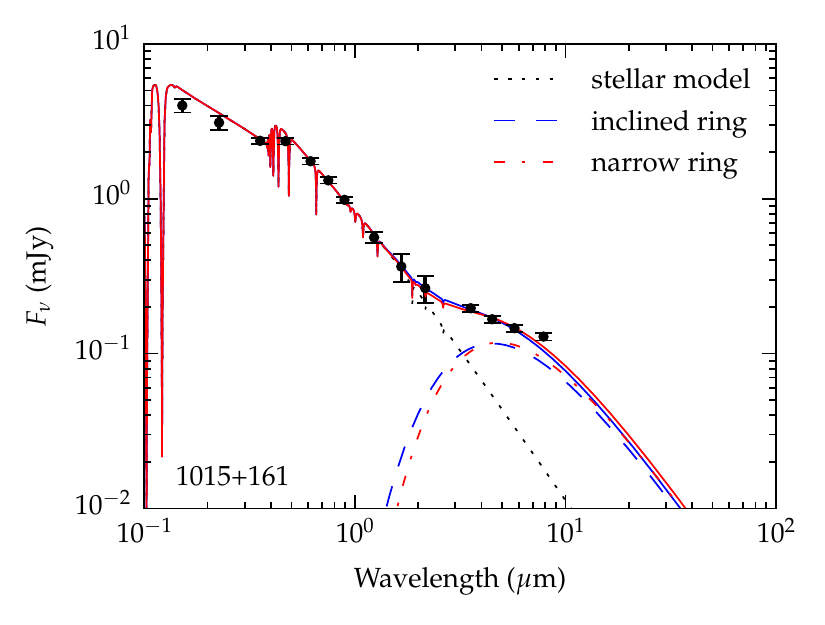}
\includegraphics{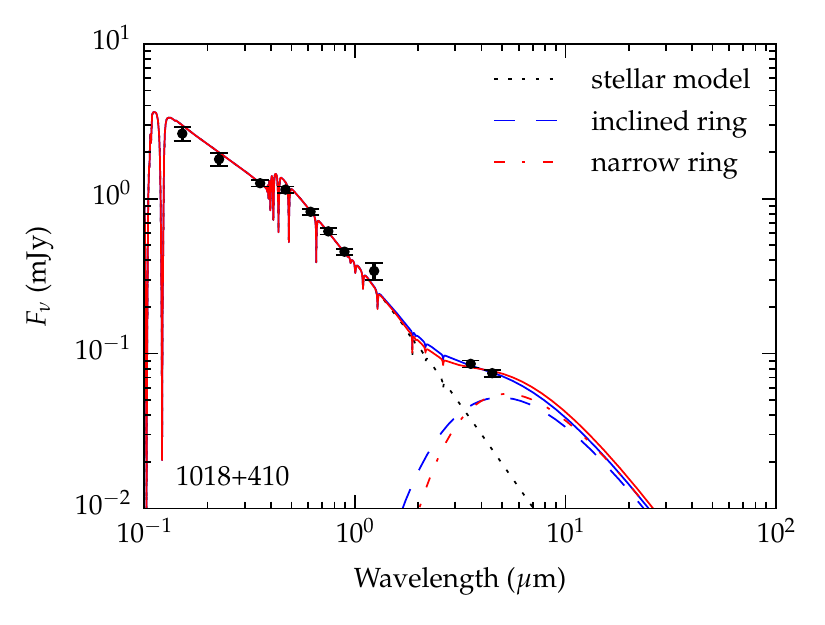}
\includegraphics{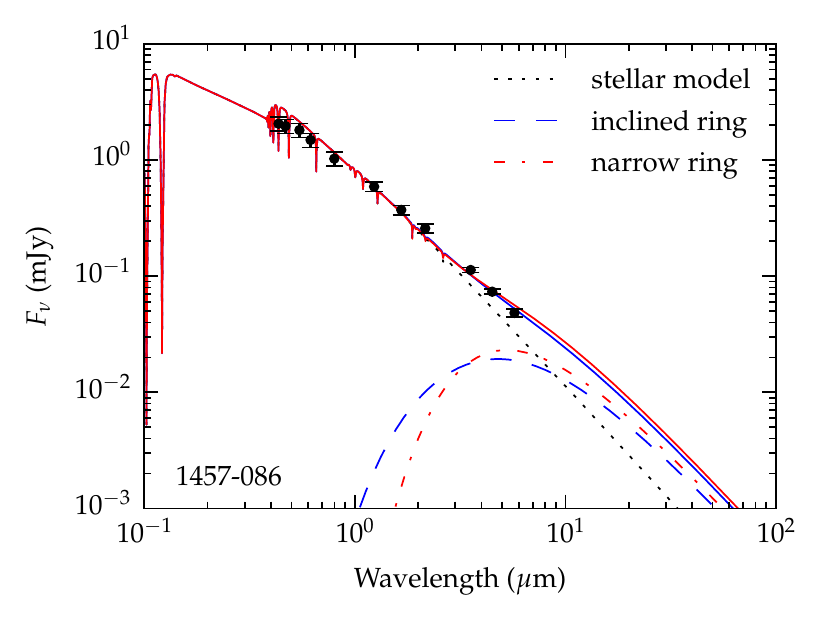}
\includegraphics{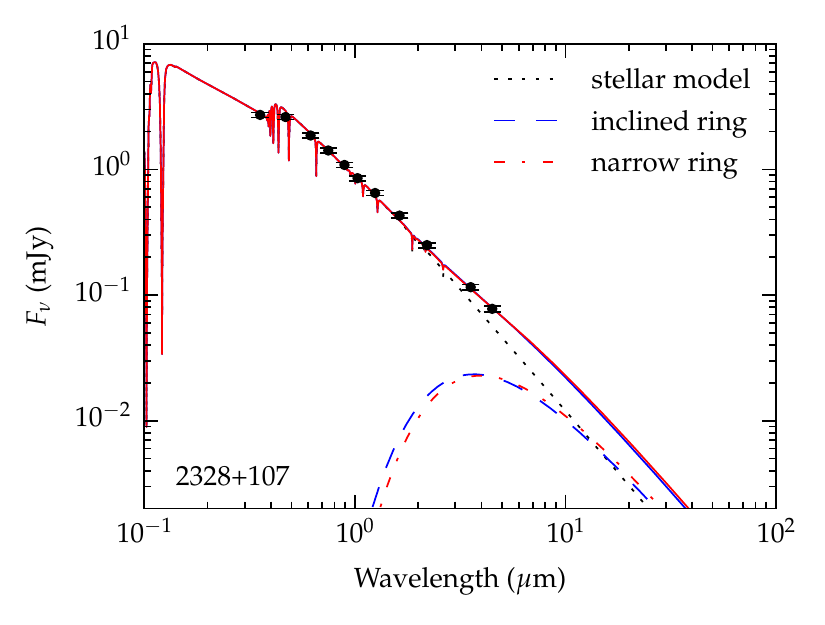}
\caption{Infrared excesses fit by circumstellar disc models, with parameters listed in Table \ref{tab:dust_discs}. Stellar atmosphere models are plotted as dot-dashed lines. Dashed lines represent emission from optically thick disc models, and solid lines represent the sum of the stellar and disc model fluxes. Two disc models are plotted for each star, one corresponding to highly inclined discs (blue), one to narrow rings (red). Circle symbols with error bars represent optical and infrared photometry, including IRAC fluxes.}
\label{fig:discmodels}
\end{figure*}

Amongst the 134 sample stars, a total of five white dwarfs show a significant ($>4\sigma$) excess in the IRAC bands: 0843+516, 1015+161,   1457--086, 1018+410,  and 2328+107. The first three stars have known infrared excesses \citep{2012ApJ...745...88X,2009ApJ...694..805F,2007ApJ...663.1285J} and are also known to be metal polluted \citep{2012MNRAS.424..333G,2005A&A...432.1025K}, while the infrared excesses at 1018+410 and  2328+107 are reported here for the first time. The presence of a significant excess at the ancillary target 1929+012 \citep{2011ApJ...732...90M,2011ApJS..197...38D,2010MNRAS.404L..40V} is also confirmed. This target, which is not part of the statistical sample, is discussed separately in Section 7, together with notes on individual excesses. The infrared excesses  were individually modelled as thermal continua using an optically thick, geometrically thin disc \citep{2003ApJ...584L..91J}. Stellar, substellar or planetary companions, and background contamination were confidently ruled out as the excess emission is either too strong \citep{2009ApJ...694..805F,2008ApJ...681.1470F} or the infrared colours are not compatible. The white dwarf radius was estimated from evolutionary models \citep{2001PASP..113..409F} while the distance was derived using synthetic absolute photometry \citep{2006AJ....132.1221H} compared with available optical and near-infrared magnitudes. Fluxes at \mbox{7.9 $\mu$m} were excluded from the fits as these are often contaminated by a strong, $10\,\mu$m silicate emission feature, as seen in e.g. GD\,362 \citep{2009AJ....137.3191J} and G29-38 \citep{2005ApJ...635L.161R}.

The free parameters in the disc model are the inner disc temperature $T_\mathrm{in}$, the outer disc temperature $T_\mathrm{out}$, and the disc inclination $i$. The radius of the disc at these temperatures can be easily estimated and is proportional to $T^{-3/4}$ \citep{1997ApJ...490..368C}. In the absence of longer wavelength photometry extending to 24\,$\mu$m, these three free parameters cannot be well constrained \citep{2007ApJ...663.1285J}. A modest degree of degeneracy is found between the inclination and the radial extent of the disc, especially for subtle excesses \citep{2014MNRAS.444.2147B,2012ApJ...749..154G}. While the inner disc temperature and radius can be well constrained by the 3.6\,$\mu$m emission, a large set of outer radii and disc inclinations can equally fit the longer-wavelength excess \citep[see e.g.][]{2007ApJ...663.1285J}. Two sets of representative models are therefore presented for each star with infrared excess, one corresponding to a relatively wide disc at high inclination and one to a more narrow ring at low inclination.  An algorithm for bound constrained minimisation \citep{Byrd:1995} was used to estimate the best fit parameters in both cases.

Figure \ref{fig:discmodels} shows the modelled and photometric SEDs of the dusty white dwarfs, and Table \ref{tab:dust_discs} gives the fitted disc parameters for the two sets of models. It can be seen that the outer radii of the model discs are all located within the  Roche limit of the star ($\approx 1.2\,R_\odot$) where planetesimals larger than $\approx1$\,km would be tidally destroyed. Moreover, the acceptable values for the inner disc temperatures are  in agreement with the temperature at which solid dust grains rapidly sublimate in a metal-rich and hydrogen-poor disc \citep{2012ApJ...760..123R}. It is interesting to notice that all models with larger radial extent require improbably high inclinations, all greater or equal to 85$^\circ$ (Table \ref{tab:dust_discs}). Because it is highly unlikely for all the inclinations to be so confined, most if not all these discs must actually be relatively narrow. Section  \ref{sec:tau} discusses this evidence in greater detail.

\begin{table}
      \caption{White dwarfs with circumstellar discs in this \emph{Spitzer} survey.}
  \begin{threeparttable}
  \begin{tabular}{@{}lcccccc@{}}
  \hline
WD	                       & $R/d$         & $T_\mathrm{in}$ & $T_\mathrm{out}$ & $r_\mathrm{in}$ & $r_\mathrm{out}$  & $ i$   \\
  	                       & ($10^{-12}$)  &  (K)              & (K)              & ($R_*$)         & ($R_*$)           & (deg)  \\ 
\hline
\multicolumn{7}{c}{\rule{0pt}{4ex}Model 1: High inclinations} \\
\rule{0pt}{4ex}0843+516	                & 3.2           & 1750            & 800             & 14              & 30                &  87      \\
1015+161                & 4.2           & 1450            & 900             & 14              & 22                &  85        \\
1018+410	                & 2.9           & 1600            & 700             & 14              & 32                &  87         \\
1457--086               & 2.8           & 2000            & 600             & 10              & 34                &  89        \\
2328+107	                & 4.4           & 1480            & 1300             & 14              & 16                &  88       \\
\multicolumn{7}{c}{\rule{0pt}{4ex}Model 2: Narrow rings} \\
\rule{0pt}{4ex}0843+516	                & 3.2           & 1000            & 960             & 24              & 25                &  30      \\
1015+161                & 4.2           & 1120            & 1000             & 18              & 20                &  70        \\
1018+410	                & 2.9           & 1000            & 940             & 22              & 24                &  40         \\
1457--086               & 2.8           & 1000            & 940             & 20              & 21                &  75        \\
2328+107	                & 4.4           & 1300            & 1270             & 16              & 17                &  75       \\
\hline
\end{tabular}
\end{threeparttable}
\label{tab:dust_discs}
\end{table}

\section{The frequency of circumstellar debris}

The detection of dust at five of 134 sample stars translates to a nominal excess frequency of  $3.7^{+2.4}_{-1.0}$\% for post-main sequence ages of 15--270\,Myr. The upper and lower bounds are calculated using the binomial probability distribution and $1\sigma$ confidence level. This frequency is a firm lower limit on the fraction of 2--3 $M_\odot$ stars (the typical progenitors of the white dwarfs in our sample) that form planetary systems.
However, complementary \emph{HST} COS observations demonstrate that at least 27\% of white dwarfs with diffusion timescales of only a few weeks have photospheric metals that require ongoing accretion of a circumstellar reservoir. These findings clearly indicate that the actual fraction of debris discs at white dwarfs is almost an order of magnitude higher, with nearly 90\% of discs emitting insufficient flux to be detected by the current infrared facilities, such as \emph{Spitzer}.

\subsection{Complementary \emph{HST} COS observations}

The entire sample of 134 stars observed with \emph{Spitzer} was  also approved as an \emph{HST} COS Snapshot program. Snapshot targets are observed during gaps between regular guest observer programs, and a total of 85 of the 134 white dwarfs were observed between 2010 September and 2013 February within programs 12169 and 12474.  The results of the  \emph{HST} survey itself are published elsewhere \citep{2014A&A...566A..34K}, and these are summarised briefly here. It was found that amongst these 85 stars, 56\% display atmospheric metals: 48 exhibit photospheric Si, 18  also show C, and 7 show further metals. The analysis indicated that for 25 stars the metal abundances may be explained by radiative levitation alone, although  accretion has likely occurred recently, leaving  23 white dwarfs (27\%) that exhibit traces of heavy elements that can only be explained with ongoing accretion of circumstellar material, in  agreement with previous estimates \citep{2010ApJ...722..725Z,2003ApJ...596..477Z}.  \emph{Spitzer} observations of the 85 star subsample with \emph{HST} data show that, amongst the 23 metal polluted white dwarfs that are currently accreting, there are two that exhibit detectable infrared excesses: 0843+516 and 1015+161. The co-observed subsample thus translates to approximately 10\% of metal-enriched stars exhibiting detectable dust, indicating that about 90\% of debris discs escape detection in the infrared. As expected, no infrared excesses are confidently found in the subsample of non-metal bearing degenerates observed with \emph{HST}, strongly supporting the connection between infrared excesses and metal pollution.

Possible reasons for the apparent lack of infrared disc detections are still a matter of debate. The collective data for the known circumstellar discs at hydrogen dominated white dwarfs indicate that the DA degenerates with the highest accretion rates are significantly more likely to host an infrared detectable circumstellar  disc \citep{2012ApJ...745...88X,2009ApJ...694..805F,2007ApJ...663.1285J}. The infrared excesses at 0843+516 and 1015+161 confirm this trend, as they both have the highest inferred Si accretion rates amongst the \emph{HST} sample stars ($3.6\times10^7$ g\,s$^{-1}$ and $5\times10^6$ g\,s$^{-1}$ respectively). This is consistent with a picture where white dwarfs accreting at the highest rates require the most massive and highest surface density discs, which are more likely to be detected in the infrared. 
One possible explanation for the dearth of infrared excesses is that mutual collisions may be enhanced in low surface density discs, and result in the partial or complete destruction of dust grains \citep{2007ApJ...663.1285J}. A related possibility is increased, collisional grain destruction due to the impact of additional, relatively small planetesimals on a pre-exiting dust disc \citep{2008AJ....135.1785J}. This may explain the lower frequency of infrared excess detections at older and cooler white dwarfs  \cite[][see also Section \ref{sec:tau}]{2014MNRAS.444.2147B}, as the depletion of the reservoir of large asteroids for older white dwarfs would imply that smaller planetesimals are primarily accreted. 
In contrast, this work corroborates the interpretation that the lack of infrared excess detections at a large fraction of metal polluted white dwarfs is largely caused by the  small total emitting surface area of dust grains, which implies a very low infrared fractional luminosity and hence undetectability with the current instrumentation. Sections \ref{sec:residuals} and \ref{sec:tau} provide evidence supporting this hypothesis.

\subsection{Hidden subtle excesses in the DAZ sample}
 
\label{sec:residuals}

Thanks to the large number of stars observed as part of this \emph{Spitzer} survey, one can investigate the possibility that some DAZ white dwarfs have an infrared excess that is just below the current sensitivity limit. The distribution of excess and deficit infrared fluxes with respect to the model fluxes demonstrates a correlation between subtle infrared excesses and atmospheric metals indicating a population of tenuous circumstellar discs.  

The observed excess and deficit fluxes with respect to the model flux at different wavelengths were expressed as a fraction of the photometric and model uncertainties, and were quantified by an excess significance, defined as:
\begin{equation}
\chi = \frac{F_\mathrm{obs} - F_\mathrm{model}}{\sqrt{\sigma_\mathrm{obs}^2+\sigma_\mathrm{model}^2}}
\end{equation} 
where  $F_\mathrm{obs}$ and $F_\mathrm{model}$ are the observed and photospheric model fluxes respectively, and similarly for the uncertainties, $\sigma_\mathrm{obs}$ and $\sigma_\mathrm{model}$. 

\begin{figure}
\includegraphics{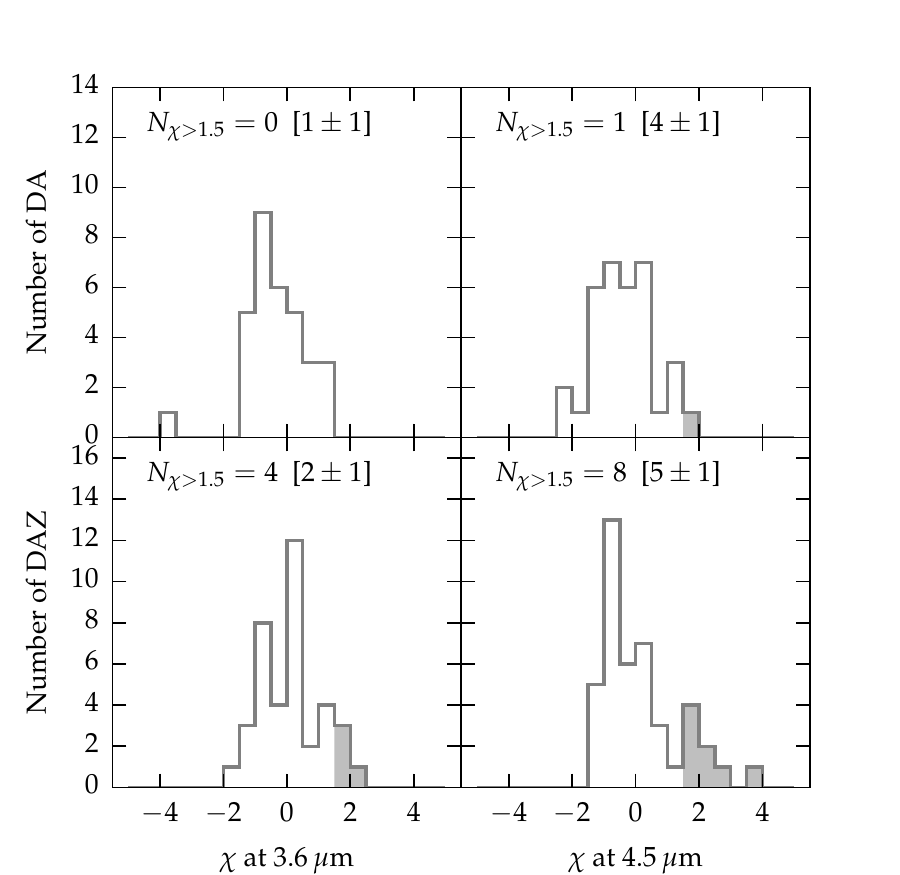}
\caption{Histograms of excess significance $\chi$ for metal lined (DAZ) and non-metal bearing (DA) white dwarfs observed at 3.6\,$\mu$m and 4.5\,$\mu$m with IRAC. Stars with $\chi > 1.5$ are highlighted in grey, and their occurrence is given at the top left of each subplot, with expected values in square brackets. Stars with $\chi > 4$ and sources that are heavily contaminated by nearby objects have been excluded.  }
\label{fig:excesssignificance}
\end{figure}

The excess significance values were derived for each star in the subsample of 85 degenerates that have both \emph{Spitzer} and \emph{HST} observations. Two subgroups were created, one containing non-metal lined (DA) white dwarfs only and one containing degenerates exhibiting atmospheric metals (DAZ) only.   All stars that show Si absorption in their spectra -- including those whose metals can be explained by radiative levitation alone -- were included in the DAZ sample, as these stars have likely accreted circumstellar material recently \citep{2014A&A...566A..34K}. The DA sample consists of 32 stars observed at 3.6\,$\mu$m and 38 stars at 4.5\,$\mu$m, while the DAZ sample consists of 34 stars observed at 3.6\,$\mu$m and 45 stars at 4.5\,$\mu$m. The differences in the subsample sizes are due to the lack of IRAC observations in channel 1 or 2 for some stars whose data were taken from the \emph{Spitzer} archive. Infrared excesses that have $\chi > 4$ and cases of strong contamination from nearby sources were excluded. The excess significance values were then plotted in separate histograms, for fluxes measured at 3.6\,$\mu$m and 4.5\,$\mu$m. The resulting four histograms are shown in Figure \ref{fig:excesssignificance}.
It is important to stress that statistically confident excesses are defined for $\chi > 4$, corresponding to a significance level of 4$\sigma$. An excess with $1.5 < \chi < 4$ can be defined as a \emph{candidate subtle excess}, which cannot yet be confirmed with confidence. However, one can test the observed distribution of all $\chi < 4$ for each subsample of DA and DAZ stars observed at 3.6\,$\mu$m and 4.5\,$\mu$m against expectations. 

The number of candidate subtle excesses with $\chi > 1.5$ and their expected values were computed for each of these subsamples, and these are shown in Figure \ref{fig:excesssignificance}. The expected values were inferred assuming no correlation between atmospheric metals and infrared excesses and were computed in the following way. The observed excesses were randomly distributed in two subsamples equal in size to the observed DA and DAZ subsamples, at  3.6 and 4.5\,$\mu$m. This computation was repeated 1000 times, and the average number of stars with $\chi > 1.5$ and its standard deviation were used as estimates. 

Interestingly, departure from expectations was found. In the DA sample there is only one candidate excess at 4.5\,$\mu$m, while the expected number is $1\pm1$ at 3.6\,$\mu$m and $4\pm1$ at 4.5\,$\mu$m. In contrast, the bulk of candidate excesses are found in the DAZ sample, where four and eight $\chi > 1.5$ values were found at 3.6\,$\mu$m and 4.5\,$\mu$m, while the expected numbers are $2\pm1$ and $5\pm1$ respectively.

This apparent correlation between atmospheric metals and subtle infrared excesses, especially at longer wavelengths, may reinforce the idea that most if not all circumstellar discs at metal polluted white dwarfs harbour dust that emits in the infrared, but its signature is too subtle to be detected with the current instrumentation. 

\section{Comparison with previous surveys}

In this section the frequency of infrared excesses at white dwarfs derived in this work is compared with previous surveys.

Implementing a $ugriz$  colour  selection valid for $T_\mathrm{eff} > 8000$\,K, \cite{2011MNRAS.417.1210G} selected 1884 DA white dwarf candidates from the SDSS DR7 catalogue spatially cross-correlated with the UKIDSS catalogue,  and found an 0.8\% of infrared bright disc candidates. This frequency includes a correction for the contamination of hot stars and sub-luminous B stars. Amongst the photometrically selected stars, about 1000 white dwarfs were further identified spectroscopically. Restricting this sample to the effective temperature range of the COS/IRAC survey it was found that for $T_\mathrm{eff} = 17\,000 - 25\,000$\,K a fraction of 6.2\% have infrared excesses. This frequency is higher than that derived from this \emph{Spitzer} survey, but includes non-discs such as low mass stellar and substellar companions and closely aligned background sources.
Similarly, \cite{2011MNRAS.416.2768S} identified 811 DA white dwarfs with $T_\mathrm{eff} < 100\,000\,$K in common between SDSS DR4, or \cite{1999ApJS..121....1M} catalogues, and UKIDSS. It was found that only three stars have a $K$-band excess consistent with a debris disc, translating to a nominal frequency of 0.4\%. Direct comparison with this work is not possible, due to the disparate effective temperature regimes and the lack of on-line data.

Mid-infrared photometry from the preliminary \emph{Wide-field Infrared Survey Explorer} \citep[\emph{WISE;}][]{2010AJ....140.1868W} catalogues, cross-correlated with the preliminary SDSS DR7 white dwarf catalogue, resulted in 1527 \emph{WISE} detections and 52 disc candidates \citep{2011ApJS..197...38D}. The frequency of discs was estimated applying a flux cutoff in the \emph{WISE} photometry, resulting in six disc candidates amongst 395 white dwarfs, and yielding a lower frequency of $1.5\% \pm 0.6\%$. Further analysis of the sample revealed that two of these candidates are spurious, so that a new estimate for the frequency of discs in the brightness limited \emph{WISE} sample is 1\% \citep{2014ApJ...786...77B}. This sample had no restriction in effective temperature and included $T_\mathrm{eff} > 25\,000\,$K stars, for which discs have never been confirmed.

These frequencies should be considered with care, as disc candidates often need to be confirmed with additional data. For instance, \emph{Spitzer} observations of six disc candidates from \cite{2011MNRAS.417.1210G} reveal that only five of the excesses are confirmed at longer wavelengths, and only three are due to dust discs \citep{2012MNRAS.421.1635F}. Prior \emph{Spitzer} observations also show that only 50\% of dusty white dwarfs show an excess in $K$-band \citep{2010ApJ...714.1386F}. 
Lastly, further follow-up of 16 disc candidates in the \emph{WISE} survey \citep{2011ApJS..197...38D} revealed that the majority of the detected excesses are due to background contamination (\citealp{2014ApJ...786...77B}; \citealp*{2014ApJ...782...20W}), showing that excess photometric fluxes detected with \emph{WISE} need additional confirmation. It is also worth noting that, although many \emph{WISE} disc candidates might be spurious, it is likely that all white dwarfs with bright dust discs have been discovered for the bulk of known white dwarfs \citep{2013ApJ...770...21H}. 

\cite{2007ApJS..171..206M} presented the first unbiased \emph{Spitzer} surveys of white dwarfs in search of infrared excesses. A brightness limited sample of 135 single stars from \cite{1999ApJS..121....1M} and with $K_s < 15$\,mag in 2MASS was selected to be observed at 4.5 and 7.9\,$\mu$m. Infrared fluxes were successfully measured for 124 of these, and only two showed significant excess due to dust, yielding a frequency of 1.6\%. The sample stars have a large temperature range, extending to hot effective temperatures at which solid dust grains rapidly sublimate within the Roche limit of the star. This brightness limited sample was naturally diverse in effective temperature, stellar mass, and atmospheric composition, making it problematic to compare with other surveys.

\cite{2012ApJ...760...26B} report a $4.3^{+2.7}_{-1.2}$\% frequency of discs at DA white dwarfs but this result contains biases due to 1) the method used to select the statistical sample and 2) the strategy to detect the discs. 
A total of 117 single white dwarfs with $T_\mathrm{eff}=9500-22\,500$\,K were selected from the Palomar Green (PG) Survey \citep*{1986ApJS...61..305G}, but the authors did not provide details on the actual selection. There are in fact 145 single degenerates in the PG catalogue in this temperature range, and all of these stars should have been considered for an unbiased frequency.
The frequency reported is also significantly higher than that found by other surveys, especially considering the 
preliminary $K$-band excess selection method. As mentioned previously, only about half of  known dusty white dwarfs have a $K$-band excess \citep{2010ApJ...714.1386F}, and  this 4.3\% frequency  represents only a lower limit. This means that excesses that show their signature beyond 2.2\,$\mu$m have likely been missed. Indeed, one of the stars in the \cite{2012ApJ...760...26B} sample, PG\,2328+108, was observed as part of this \emph{Spitzer} survey, and a subtle excess was confidently detected (see Figure \ref{fig:discmodels}).
It is also worth noting that 2MASS photometry becomes unreliable for sources fainter than $K_s \approx 14$\,mag  \citep{2009MNRAS.398.2091F,2007AJ....134...26H}  and most of their 39 stars with only 2MASS photometry are fainter than this limit.
Finally, four previously known dusty white dwarfs (1015+161, 1116+026, 1457--086 and 2326+049) were included in the sample of 117 stars, but were not followed-up photometrically or spectroscopically, and were assumed to have discs a priori. This can introduce a strong bias in disc frequency, as two of these discs (1015+161,  1116+026) do not actually have detectable $K$-band excess ($J-K = 0.01 \pm 0.10$ and $J-K = 0.08 \pm 0.10$ respectively) and would have been missed if not previously known.

Despite these issues, it is worth trying to derive a robust lower limit for the frequency of detectable discs in the PG catalogue. Considering the infrared excess discovered at 2328+108, together with the other five known infrared excesses detected in the \emph{complete} PG sample of 145 stars with $T_\mathrm{eff}=9500-22\,500$\,K, a lower limit for the disc frequency in this temperature range is  4.1\%.  Note that other subtle infrared excesses might still be present, as not all 145 stars have \emph{Spitzer} observations. 
Restricting the range of effective temperatures to $T_\mathrm{eff}=17\,000-22\,500$\,K allows direct comparison with this work. There are 96 stars and five infrared excesses in this temperature range in the sample analysed in this study, translating to a 5.2\% frequency. Similarly, there are three infrared excesses amongst 50 stars in the PG sample with $T_\mathrm{eff}=17\,000-22\,500$\,K, yielding a frequency of $\ge 6\%$. For comparison, \cite{2011MNRAS.417.1210G} found a higher fraction of 6.9\% $K$-band excesses in this temperature range, but this frequency does not take into account contaminant sources.  

\section{The fractional luminosity of dust discs}
\label{sec:tau}

\begin{figure*}
\vspace*{1.5cm}
\includegraphics{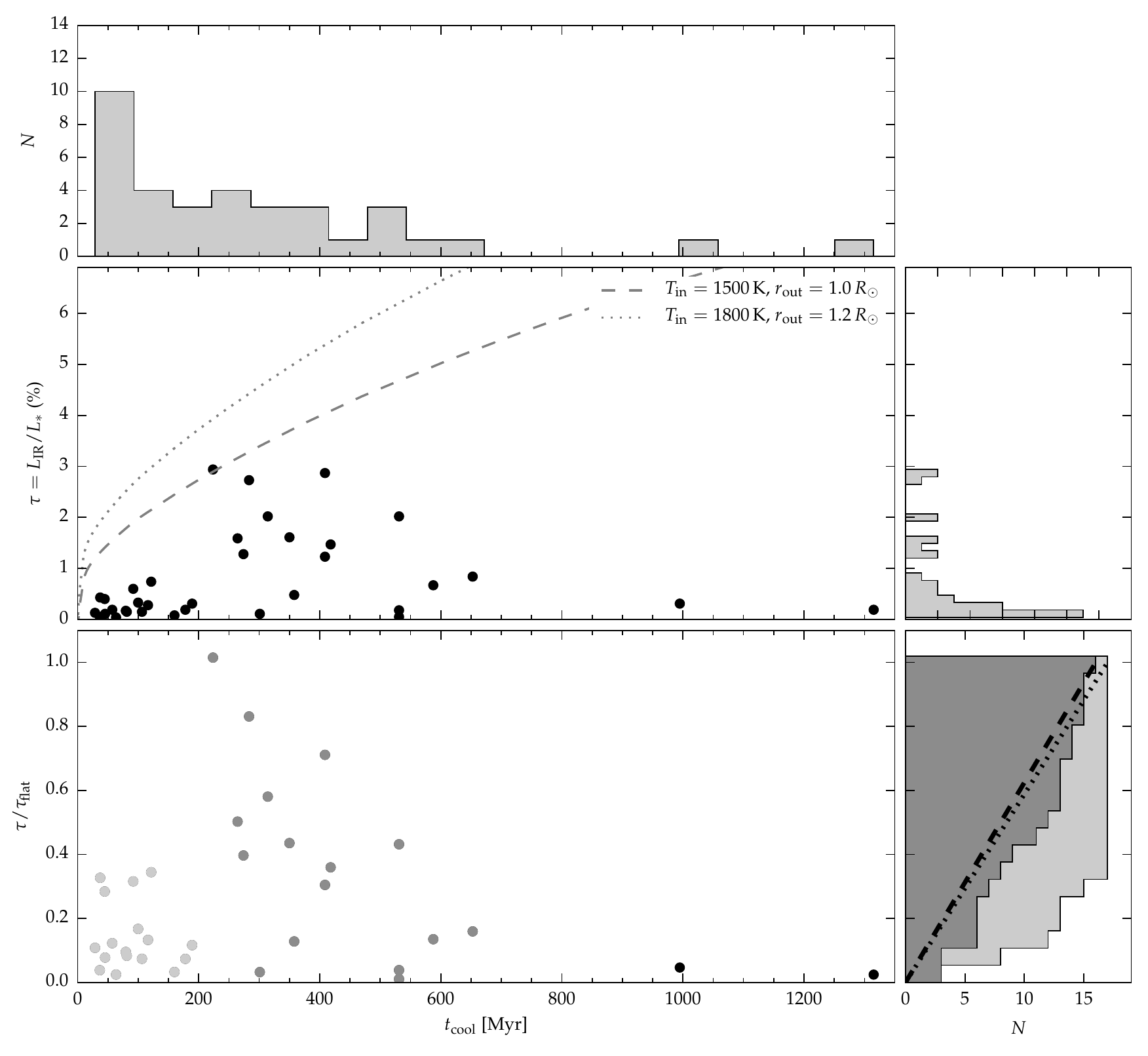}
\caption{Fractional disc luminosities ($\tau = L_\mathrm{IR}/L_\mathrm{*}$) for all known white dwarfs with detected dust discs. The central panel shows the value of $\tau$ as a function of cooling age for $\log g = 8$. The two lines lines show the predicted  fractional luminosity for a face on, opaque flat disc with radial extent given by assuming dust grains persist up to 1500 and 1800\,K and extend up to 1.0\,$R_\odot$ and 1.2\,$R_\odot$. The dotted line is for  $T_\mathrm{in} = 1500$\,K and $r_\mathrm{out} = 1.0\,R_\odot$, while the dashed line is for $T_\mathrm{in} = 1800$\,K and $r_\mathrm{out} = 1.2\,R_\odot$. 
The top panel shows the histogram of the number of infrared excesses as a function of cooling age, while the right panel shows the histogram of $\tau$ with 20 bins between 0 and 3\%.
The bottom panel shows the ratio between the infrared fractional luminosity of all known dusty white dwarfs and the predicted maximum fractional luminosity for a flat, passive disc, assuming $T_\mathrm{in} = 1500\,$K and $r_\mathrm{out} = 1.0\,R_\odot$. The light grey symbols have $t_\mathrm{cool} < 200\,\mathrm{Myr}$, while the dark grey symbols have $200\,\mathrm{Myr} < t_\mathrm{cool} < 700\,\mathrm{Myr}$.  The bottom-right panel shows the cumulative histograms of $\tau/\tau_\mathrm{flat}$ for the light and dark grey target subsets, and the expected distributions for random inclinations are shown as dotted and dashed lines respectively.}
\label{fig:tau}
\end{figure*}

\begin{table}
  \caption{Estimated fractional luminosities from thermal continuum for all known white dwarfs exhibiting infrared excess due to dust. }
  \begin{threeparttable}
  \begin{tabular}{lrrrcr}
  \hline
WD & $ t_\mathrm{cool} $ & $T_\mathrm{eff}$ & $T_\mathrm{IR}$ & $\tau$  & Ref. \\
   &     (Myr)      &    (K)             &        (K)      &   (\%)  &      \\
\hline
0106--328  &  160  &  16000  &   1470  &  0.08&  1  \\
0110--565  &   81  &  19200  &   1050  &  0.15&  2  \\
0146+187  &  418  &  11500  &   1120  &  1.47&  3  \\
0246+734\tnote{\textdagger}  &  995  &   8250  &   1000  &  0.31&  4  \\
0300--013  &  189  &  15200  &   1200  &  0.31&  5  \\
0307+078  &  531  &  10500  &   1200  &  0.18&  1  \\
0408--041  &  224  &  14400  &   1020  &  2.94&  5  \\
0435+410  &  116  &  17500  &   1250  &  0.28&  2  \\
0735+187  &  264  &  13600  &   1330  &  1.59&  6  \\
0842+231  &   92  &  18600  &   1350  &  0.60&  6  \\
0843+516  &   29  &  23900  &   1310  &  0.13&  7  \\
0956--017  &  283  &  13280  &   1220  &  2.73&  8  \\
1015+161  &   80  &  19300  &   1210  &  0.17&  5  \\
1018+410  &   45  &  22390  &   1210  &  0.11&  9  \\
1041+091  &  106  &  17910  &   1500  &  0.15&  6  \\
1116+026  &  358  &  12200  &   1000  &  0.48&  5  \\
1150--153  &  314  &  12800  &    940  &  2.02&  10  \\
1219+130  &  350  &  12300  &   1250  &  1.61&  8  \\
1225--079  &  531  &  10500  &    300  &  0.05&  1  \\
1226+110  &   45  &  22000  &   1070  &  0.40&  11  \\
1349--230  &  100  &  18200  &   1260  &  0.33&  2  \\
1455+298  & 1315  &   7400  &    400  &  0.19&  12  \\
1457--086  &   63  &  20400  &   1400  &  0.04&  3  \\
1541+651  &  409  &  11600  &    980  &  1.23&  13  \\
1551+175  &  178  &  15500  &   1690  &  0.19&  4  \\
1554+094  &   37  &  22800  &   1100  &  0.43&  8  \\
1615+164  &  274  &  13430  &   1370  &  1.28&  6  \\
1729+371  &  531  &  10500  &    820  &  2.02&  14  \\
1929+011  &   57  &  20890  &   1060  &  0.19&  5  \\
2115--560  &  653  &   9700  &    820  &  0.84&  3  \\
2132+096 &  301  &  13000  &    550  &  0.11&  4  \\
2207+121  &  122  &  17300  &   1100  &  0.74&  7  \\
2221--165  &  588  &  10100  &    950  &  0.67&  1  \\
2326+049  &  409  &  11600  &   1060  &  2.87&  12  \\
2328+107  &   37  &  21000  &   2000  &  0.05&  9  \\
 \hline
  \end{tabular}
\begin{tablenotes}
 \small
\item[\textdagger]  Needs to be confirmed.
  \item \textbf{References:} \\
(1) \cite{2010ApJ...714.1386F}
(2) \cite{2012ApJ...749..154G}
(3) \cite{2009ApJ...694..805F}
(4) \cite{2014MNRAS.444.2147B} 
(5) \cite{2007ApJ...663.1285J} 
(6) \cite{2012ApJ...750...86B}
(7) \cite{2012ApJ...745...88X}
(8) \cite{2003ApJ...596..477Z,2012MNRAS.421.1635F}
(9)  this work; 
(10) \cite{2009AJ....137.3191J}
(11) \cite{2009ApJ...696.1402B}
(12) \cite{2008ApJ...674..431F}
(13) \cite{2012ApJ...760...26B}
(14) \cite{2007ApJ...663.1285J}
\end{tablenotes}
\end{threeparttable}
\label{tab:tau}
\end{table}

Here it is shown that the distribution of the disc fractional luminosities (\mbox{$\tau = L_\mathrm{IR}/L_*$}) for the sample of known dusty white dwarfs reinforces the interpretation that a large fraction of circumstellar discs remain undetected in the infrared and may also provide  insight into their evolution.

The value of fractional luminosity of all known infrared excesses at cool white dwarfs was estimated in a consistent manner. The photospheric flux was fitted as described in Section \ref{sect:data-analysis} and a blackbody distribution was then fitted to the infrared excesses seen in the $K$-band and the three shortest wavelength IRAC bandpasses. The mid-infrared fluxes were all measured with \emph{Spitzer} IRAC and were obtained from the literature and this work. Although a precise estimate of $\tau$ requires detailed modelling of the photospheric and disc flux, the values reported here are estimated to be good to 10\%. Independently measured values for 18 discs known as of mid-2010 agree to within 10\% \citep{2011wdac.book..117F}. Note also that $\tau$ is the \emph{observed} fractional luminosity and it does not take into account the disc inclination, which cannot be determined confidently (Section \ref{sec:infrared_excesses}). The fractional luminosity $\tau$ and the blackbody dust temperature $T_\mathrm{IR}$ for all 35 \emph{Spitzer} detected white dwarfs with infrared excesses are listed in Table \ref{tab:tau}. The measured values of $\tau$ and their distribution are also plotted in Figure \ref{fig:tau}.

\subsection{An undetected population of subtle excesses}
The right hand histogram of the middle panel of Figure \ref{fig:tau} shows the distribution of $\tau$, with 20 bins between 0\% and 3\%. It can be seen that the distribution rises sharply at the smallest fractional luminosities indicating that the majority of circumstellar discs are subtle. This suggests that many discs might have a fractional luminosity that is below the sensitivity limits of the current instrumentation and therefore escape detection. 

Together with the likely presence of several additional subtle excesses seen in this \emph{Spitzer} sample (see Section \ref{sec:residuals}), the distribution of $\tau$ points to a large population of subtle excesses, suggesting  that most if not all, currently accreting white dwarfs with metals have circumstellar dust.  This significant population of dust discs are not detectable with current facilities, likely due to low surface areas and optical depth.

\subsection{Narrow rings at young white dwarfs}

The central panel of Figure \ref{fig:tau} shows $\tau$ as a function of cooling age. The two lines represent the  value of the maximum fractional disc luminosity $\tau_\mathrm{flat}$ for a flat disc model \citep{2003ApJ...584L..91J} with a face-on configuration. This value was calculated assuming that the dust occupies all the space available between the distance from the star at which dust grains rapidly sublimate and the stellar Roche limit. The two lines correspond to two different assumptions about the maximum disc extent. The dotted line assumes that silicates rapidly sublimate at $T_\mathrm{in} = 1800\,$K and the stellar Roche limit is $r_\mathrm{out} = 1.2\,R_\odot$, while the dashed line assumes $T_\mathrm{in} = 1500\,$K and $r_\mathrm{out} = 1.0\,R_\odot$. 

From the central panel it can be seen that infrared excesses have so far only been detected at  white dwarfs older than \mbox{$\approx 25\,$Myr}, and with corresponding $T_\mathrm{eff} \lesssim 25\,000\,$K. This is somewhat expected, as at higher stellar temperatures, any disrupted asteroids debris will evaporate relatively quickly, so that their detection is less likely. However, the calculations done here and shown in the plot indicate that optically thick discs may exist at higher effective temperatures and cooling ages. Such discs would likely evolve more rapidly  \citep{2011MNRAS.416L..55R}, and have a higher gas to dust ratios from sublimation prior to settling into optically thick rings. 

As the white dwarf cools below 25\,000\,K the distance at which silicates rapidly sublimate decreases. Hence, the area available for dust grains increases, so that the disc luminosity should increase as well. {\em However, the expected increase assuming that discs are fully extended (shown by the two lines in the central panel of Figure \ref{fig:tau}), is not seen in the observed fractional luminosities}. Between 25\,Myr and $200\,$Myr there are 17 dusty white dwarfs, all with fractional luminosities significantly less than the maximum value allowed for flat discs. This suggests that, at these young white dwarfs, large and extended discs do not form, or only persist for timescales significantly shorter than the disc lifetime. 

This finding can be better explored and tested in the bottom-left panel of Figure \ref{fig:tau}, showing the ratio $\tau/\tau_\mathrm{flat}$ for $T_\mathrm{in} = 1500\,$K and $r_\mathrm{out} = 1.0\,R_\odot$ as a function of cooling age. Supposing that discs are fully extended, and assuming random inclinations, the distribution of $\tau/\tau_\mathrm{flat}$ should be uniform between zero and one. However, this does not seem to be true, especially for white dwarfs younger than 200\,Myr and older than 700\,Myr. To better explore if discs have different radial extents at different ages it is useful to consider  three cooling age ranges separately, for $t_\mathrm{cool} < 200$\,Myr (light grey symbols), for $200\,\mathrm{Myr} < t_\mathrm{cool} < 700\,\mathrm{Myr}$ (dark grey symbols), and for $t_\mathrm{cool} > 700\,\mathrm{Myr}$. It can be seen that:

\begin{itemize}
\item[1)] At $t_\mathrm{cool} < 200$\,Myr, the $\tau/\tau_\mathrm{flat}$ values are all below 0.4, clearly demonstrating that inclination is not the only parameter influencing the distribution. A Kolmogorov--Smirnov (KS) test confirms this, resulting in a probability of $2 \times 10^{-7}$ that the observed distribution is uniform in $\tau/\tau_\mathrm{flat}$. As any observational bias would favour the detection of discs with lower inclinations and correspondingly higher fractional luminosities, this indicates that the observed discs must be relatively narrow at these young cooling ages.

\item[2)] Between $200\,\mathrm{Myr} < t_\mathrm{cool} < 700\,\mathrm{Myr}$ there is a notable increase in the observed fractional luminosities, with the brightest ($\tau \approx 3\%$) disc modelled as face on at 0408--041 (GD\,56). In this range of cooling ages, a KS test reveals that the distribution of $\tau/\tau_\mathrm{flat}$ is uniform with a probability of only 0.07, again indicating that a second parameter shapes the distribution, namely the disc radial extent.  While not as striking as for the youngest cooling ages, this group likely contains several relatively narrow discs, as indicated by the cumulative distribution at smaller $\tau/\tau_\mathrm{flat}$ values.

\item[3)] Lastly, at $t_\mathrm{cool} > 700$\,Myr there are only two detected discs with relatively low fractional luminosities ($\tau < 0.3\%, \tau/\tau_\mathrm{flat} < 0.05$). The small $\tau$ values of these discs, together with the decreasing frequency of detections seen at older white dwarfs (top histogram of Figure \ref{fig:tau}), suggest that the number of large asteroid disruptions per time bin may be decreasing and hence also the fraction of detectable infrared excesses \citep{2014MNRAS.444.2147B}.  However, it is interesting to note that the inferred metal accretion rates do not show a decreasing trend with cooling age \citep{2014A&A...566A..34K}, indicating that circumstellar material is present at older stars. This could be explained by the accretion of smaller asteroids, which can still provide continuous accretion, but are less likely to produce detectable amounts of dust \citep{2014MNRAS.439.3371W}.
\end{itemize}

\subsection{Possible disc evolution scenarios}

Interestingly, the existence of narrow rings is predicted by global models of white dwarf disc evolution.  In general, the outer radii of flat and optically thick discs will rapidly decrease due to Poynting-Robertson (PR) drag.  Specifically, for a range of initial, realistic surface density distributions, PR drag is significantly more efficient per unit mass on the outermost (and innermost) disc regions where grain density and optical depth is lowest.  Solids are quickly gathered inward until they result in a region of moderate optical depth (\citealp*{2012MNRAS.423..505M}; \citealp{2011ApJ...741...36B}), giving rise to a sharp outer edge.

This edge forms rapidly, and marches appreciably inward within a few to several hundred PR drag timescales \citep{2011ApJ...741...36B}, which can be as short as years for typical dusty white dwarfs and 1\,$\mu$m dust at 1\,$R_{\odot}$.  The rate at which the outer edge migrates inward will be ultimately set by the dominant grain size, as the rate scales linearly with particle size and can be a factor of 1000 longer for centimetre vs. micron sizes.  Stellar luminosity plays a smaller role, varying by less than a factor of 20 between typical 10\,000\,K and 20\,000\,K white dwarfs \citep{2001PASP..113..409F}.  Thus, if this mechanism is responsible for sculpting narrow rings, their radial extents as a function of white dwarf cooling age and luminosity may broadly constrain typical particle sizes.

The transition in disc brightness between the $t_\mathrm{cool} < 200$\,Myr and $200\,\mathrm{Myr} < t_\mathrm{cool} < 700\,\mathrm{Myr}$ cooling age ranges appears relatively abrupt, however, and it is not clear that a change in stellar luminosity (and PR drag timescale) alone can account for the difference between the two populations. It is perhaps tempting to interpret the brighter and presumably larger discs at $200\,\mathrm{Myr} < t_\mathrm{cool} < 700\,\mathrm{Myr}$ as evidence of disc spreading, but the viscous timescale among particulates, for any reasonable particle size, is orders of magnitude insufficient and thus highly unlikely to have an effect on disc evolution \citep{2012MNRAS.423..505M,2008ApJ...674..431F}.

A distinct possibility for the narrowing of rings is disc truncation by impact.  Theoretical models show that post-main sequence dynamical instabilities arise near $\sim10\,$Myr, then peak around $\sim100$\,Myr, decreasing towards later times (\citealp{2013MNRAS.431.1686V}; \citealp*{2014MNRAS.437.1404M}).  This suggests that impacts on pre-existing discs by additional, perturbed and tidally shredded asteroids would be most prevalent among the $t_\mathrm{cool} < 200$\,Myr disc sample.  Disc impacts would have the favourable characteristic that they should occur preferentially at larger orbital radii, thus typically creating outwardly truncated discs.  The impactor fragments would destroy dust masses comparable to their initial, intact masses, creating gas that would quickly dissipate \citep{2008AJ....135.1785J}.  Interestingly, the recently observed drop in infrared luminosity from the dusty white dwarf SDSS\,J095904.69$-$020047.6, and the disappearance of metallic gas emission lines from the similarly dusty and polluted star SDSS\,J161717.04+162022.4, may represent such impact events \citep{2014ApJ...792L..39X,2014MNRAS.445.1878W}.

Because viscous spreading is inefficient (see above), discs will not grow outward significantly during their lifetime, and the narrower discs might be the result of initial conditions; specifically, smaller disrupted parent bodies.  However, assuming the same overall disc properties, flat configurations and random inclinations, the average disc mass should correlate with $\tau$ over each population, implying parent body masses all similar to within a factor of roughly 50.  This would require relatively fine-tuned mass influxes per unit time in the post-main sequence, including a subtle change around 300\,Myr to include slightly larger disruptions.  Taken together, this seems unlikely to account for the disc brightness differences among the two cooling age ranges.

In summary, while none of the above mechanisms appears to be without possible drawbacks, the clear observational difference in disc populations demands a closer look with formation and evolutionary modelling.  Insufficient data could be masking a more gradual change in the fractional disc luminosities that would favour the edge migration scenario, for example. The detection of a larger number of discs, especially those producing subtle excesses, would improve the statistics and reduce the possibilities.

\section{Notes on individual targets}

\label{sec:targets}

\subsection{Sample stars with infrared excesses}

\emph{0843+516 and 1015+161}. These are two metal polluted white dwarfs  \citep{2012MNRAS.424..333G} with known infrared excesses \citep{2012ApJ...745...88X,2007ApJ...663.1285J}. The excesses at 7.9 $\mu$m lay significantly above the predicted fluxes of both disc models (Figure \ref{fig:discmodels}), likely due to the presence of a strong silicate emission feature. 

\emph{1018+410}. The infrared excess detected at this star is reported here for the first time. There is no reliable near-infrared photometry available for this star, and the 2MASS catalogue only reports $H = 16.7 \pm 0.1$ mag. However, $ugriz$ photometry is available from SDSS and the photospheric level of the white dwarf can be well constrained. This degenerate was not observed in the complementary \emph{HST} survey, and there are no high-resolution optical spectra available. Follow-up should confirm that this is a DAZ white dwarf.

\emph{1457--086}. This is a known metal polluted white dwarf \citep{2005A&A...432.1025K} that was found to host a narrow dust ring with high inclination and radial extent of only 0.01\,$R_\odot$ \citep{2009ApJ...694..805F}. The fit to the excess emission with both a narrow and highly inclined disc model show that the optical fluxes might be overestimated and 15\% error bars were assumed to obtain a satisfactory fit. The star was not observed in the complementary \emph{HST} survey and additional optical photometry may better constrain the photospheric flux.  

\emph{2328+107}. With a fractional luminosity of only 0.1\% this infrared excess, reported here for the first time, joins an expanding list of attenuated discs \citep{2014MNRAS.444.2147B,2010ApJ...714.1386F}. Both $JHK$ and $ugriz$ photometry are available from the 2MASS and SDSS online catalogues  and allow to constrain  the photospheric level confidently. This degenerate was not observed in the \emph{HST} survey and high-resolution spectra are not yet available but should reveal the presence of metals.

\subsection{The infrared excess at 1929+012}

\label{sec:1929}

\begin{figure}
\includegraphics{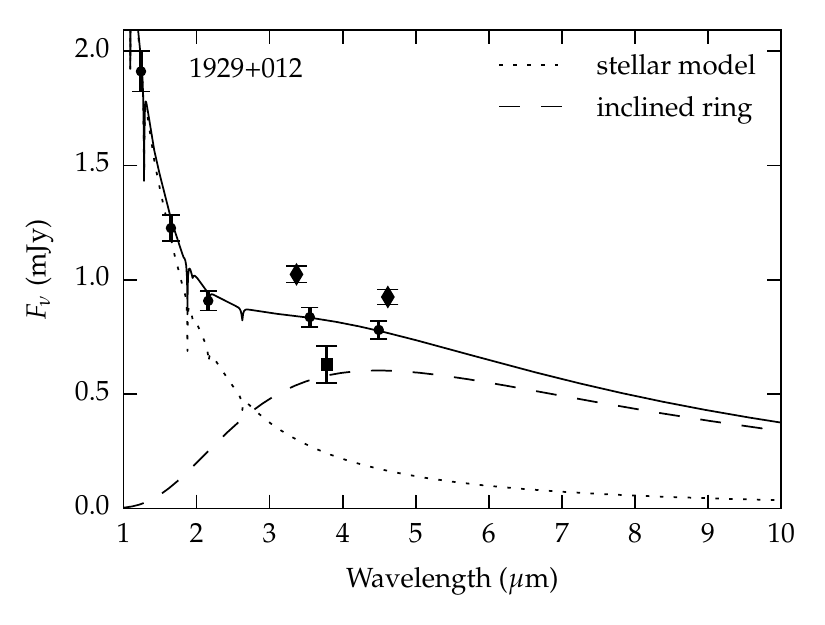}
\caption{SED and disc model for the ancillary target 1929+012. Fluxes from \emph{WISE} All-Sky catalogue are plotted as diamond symbols and fluxes from ISAAC $L'$ as square symbols. Circles symbols represent 2MASS and IRAC fluxes.}
\label{fig:GALEX1931}
\end{figure}

This star, also known as GALEX\,1931, is a metal polluted white dwarf \citep{2010MNRAS.404L..40V} with a known infrared excess \citep{2012MNRAS.424..333G,2011ApJ...732...90M,2011ApJ...729....4D}, included in the \emph{Spitzer} Program as an ancillary target. Photometry from \emph{WISE}  revealed infrared excess consistent with emission from a dust disc \citep{2011ApJ...729....4D}, while ground-based photometry out to $L$-band indicated significant contamination from background sources in the \emph{WISE} data \citep{2011ApJ...732...90M}. The \emph{Spitzer} IRAC 3.6 and 4.5\,$\mu$m photometry of GALEX\,1931 confirms the excess emission but is heavily blended with two nearby sources, as shown by the $J_sHK_sL'$ higher resolution images \citep{2011ApJ...732...90M}.  Despite the success in extracting most of the flux, the relatively low spatial resolution of IRAC implies likely contamination in the derived fluxes. Therefore, 10\% error bars are conservatively assumed in the resulting photometry. 
Figure \ref{fig:GALEX1931} shows the SED of 1929+012, including IRAC, ISAAC $L'$ \citep{2011ApJ...732...90M}, and \emph{WISE} All-Sky photometry. A fiducial disc model is also plotted. It can be seen that the IRAC fluxes are somewhat lower than the \emph{WISE} values, but also higher than the ground-based $L$-band flux.  These data indicate mild contamination is likely in the IRAC data and strong contamination is present in \emph{WISE}.

\section{Summary and conclusion}

\emph{Spitzer} observations of an unbiased sample of 134 DA white dwarfs yield a disc frequency near 4\% in the effective temperature range $T_\mathrm{eff} = 17\,000\,\mathrm{K} - 25\,000\,\mathrm{K}$. Complementary \emph{HST} observations of 85 sample stars reveal that a much larger fraction of at least 27\% host circumstellar material, indicating that about 90\% of the discs remain undetected in the infrared.

Possible reasons for the lack of infrared detections were investigated. The distribution of observed fluxes compared to photospheric models at 3.6 and 4.5\,$\mu$m for the subsamples with and without metals point to a population of excesses too subtle to currently confirm. Future observations, for example with \emph{JWST} using spectroscopy, have the potential to confirm the presence of faint dust discs. 

The distribution of the fractional disc luminosities of all known dusty white dwarfs also points towards a hidden population of subtle discs. In addition, this distribution indicates that the disc population evolves over time. Only relatively narrow rings are found at $t_\mathrm{cool} \lesssim 200$\,Myr, while relatively extended discs, filling the majority of the space available between the distance at which silicates rapidly sublimate and the stellar Roche limit, occur only after a few hundred Myr. A marked decrease in the observed  fractional disc luminosities, as well as a decrease in the frequency of detections towards cooler ages, is seen after this peak, suggesting that the number of large asteroids might be gradually depleted.

This study strongly reinforces the hypothesis that most if not all metal-enriched white dwarfs harbour circumstellar dust, but  the majority remain unseen due to low surface area and the sensitivity limits of current instrumentation.

\section*{Acknowledgments}
The authors thank the anonymous referee for suggestions that improved the manuscript.
The authors acknowledge Tom Marsh and Mark Hollands for discussions on disc geometry, and Giuseppe Morello and Ingo Waldmann for discussions on statistics.
This work is based on observations made with the \emph{Spitzer Space Telescope}, which is operated by the Jet Propulsion Laboratory, Caltech, under NASA contracts 1407 and 960785. The following databases were used in the process of this research: 
the Sloan Digital Sky Survey, which is managed by the Astrophysical Research Consortium for the Participating Institutions (http://www.sdss.org/);
the Two Micron All Sky Survey, which is a joint project of the University of Massachusetts and IPAC/Caltech, funded by NASA and the NSF;
the Deep Near-Infrared Survey of the southern sky, which has been partly funded by the SCIENCE and the HCM plans of the European Commission;
the AAVSO Photometric All-Sky Survey, funded by the Robert Martin Ayers Sciences Fund.
This research has made use of the VizieR catalogue access tool and the SIMBAD database operated at CDS, Strasbourg, France.
MR, JF, and CB collectively acknowledge support from the STFC; JF via an Ernest Rutherford Fellowship.
The research leading to these results has received funding from the European Research Council under the EU 7th Framework Programme (FP/2007--2013)/ERC Grant Agreement n. 320964 (WDTracer).
\bibliography{paper}
\newpage
\appendix 
\onecolumn
\section[]{Supplementary material}

\footnotesize
\begin{longtable}{lccccccccccc}
  \caption{Stellar parameters and flux determinations  for the science targets} \\
  \hline
WD & $T_\mathrm{eff}$ & $\log g$ & $V$ & $M_\mathrm{wd\,}$ & $M_\mathrm{ms}$ & $\log(t_\mathrm{cool})$ & $F_{3.6\,\mu\mathrm{m}}$ & $F_{4.5\,\mu\mathrm{m}}$ & $F_{5.7\,\mu\mathrm{m}}$ & $F_{7.9\,\mu\mathrm{m}}$  \\
 & (K) & [log(cm\,s$^{-2}$)] & (mag) & $(M_\odot)$ &  $(M_\odot)$ & [$\log$\,(yr)] & ($\mu$Jy) & ($\mu$Jy) & ($\mu$Jy) & ($\mu$Jy) \\
    \hline
    \endfirsthead 
    \multicolumn{3}{@{}l}{}\\
    \hline
WD & $T_\mathrm{eff}$ & $\log g$ & $V$ & $M_\mathrm{wd\,}$ & $M_\mathrm{ms}$ & $\log(t_\mathrm{cool})$ & $F_{3.6\,\mu\mathrm{m}}$ & $F_{4.5\,\mu\mathrm{m}}$ & $F_{5.7\,\mu\mathrm{m}}$ & $F_{7.9\,\mu\mathrm{m}}$  \\
 & (K) & [log(cm\,s$^{-2}$)] & (mag) & $(M_\odot)$ &  $(M_\odot)$ & [$\log$\,(yr)] & ($\mu$Jy) & ($\mu$Jy) & ($\mu$Jy) & ($\mu$Jy) \\
\hline
\endhead 
0000+171 & 20\,210 & 7.99 & 15.81 & 0.62 &  2.07 & 7.81 & $64 \pm 3$ & $44 \pm 2$ & $...$ & $...$ \\
0013--241 & 18\,530 & 7.90 & 15.38 & 0.56 &  1.56 & 7.90 & $103 \pm 5$ & $66 \pm 3$ & $...$ & $...$ \\
0018--339 & 20\,630 & 7.84 & 14.64 & 0.54 &  1.36 & 7.65 & $199 \pm 10$ & $119 \pm 6$ & $...$ & $...$ \\
0028--474 & 17\,390 & 7.65 & 15.15 & 0.44 &  ... & 7.87 & $166 \pm 8$ & $104 \pm 5$ & $...$ & $...$ \\
0047--524 & 18\,810 & 7.73 & 14.23 & 0.48 &  0.83 & 7.77 & $...$ & $189 \pm 9$ & $...$ & $65 \pm 3$ \\
0048+202 & 20\,360 & 7.89 & 15.38 & 0.57 &  1.57 & 7.72 & $101 \pm 5$ & $64 \pm 3$ & $...$ & $...$ \\
0048--544 & 17\,870 & 7.98 & 15.16 & 0.61 &  1.94 & 8.02 & $125 \pm 6$ & $81 \pm 4$ & $...$ & $...$ \\
0059+257 & 21\,370 & 8.04 & 15.90 & 0.65 &  2.35 & 7.75 & $68 \pm 3$ & $43 \pm 2$ & $...$ & $...$ \\
0102+095 & 24\,770 & 7.93 & 14.44 & 0.60 &  1.86 & 7.28 & $217 \pm 10$ & $146 \pm 7$ & $...$ & $...$ \\
0110--139 & 24\,690 & 7.99 & 15.75 & 0.63 &  2.16 & 7.36 & $60 \pm 3$ & $38 \pm 2$ & $...$ & $...$ \\
0114--605 & 24\,690 & 7.75 & 15.11 & 0.51 &  1.08 & 7.17 & $112 \pm 5$ & $75 \pm 3$ & $...$ & $...$ \\
0124--257 & 23\,040 & 7.79 & 16.18 & 0.52 &  1.18 & 7.35 & $36 \pm 2$ & $25 \pm 1$ & $...$ & $...$ \\
0127+270 & 24\,870 & 7.83 & 15.90 & 0.55 &  1.40 & 7.18 & $45 \pm 2$ & $28 \pm 1$ & $...$ & $...$ \\
0129--205 & 19\,950 & 7.88 & 15.30 & 0.56 &  1.54 & 7.75 & $103 \pm 5$ & $68 \pm 3$ & $...$ & $...$ \\
0140--392 & 21\,810 & 7.92 & 14.35 & 0.58 &  1.74 & 7.59 & $254 \pm 12$ & $165 \pm 8$ & $...$ & $...$ \\
0155+069 & 22\,010 & 7.67 & 15.47 & 0.47 &  0.66 & 7.39 & $108 \pm 5$ & $68 \pm 3$ & $...$ & $...$ \\
0200+248 & 23\,280 & 7.86 & 15.71 & 0.56 &  1.50 & 7.37 & $105 \pm 11$ & $71 \pm 7$ & $...$ & $...$ \\
0201--052 & 24\,630 & 7.64 & ... & 0.46 &  0.62 & 7.19 & $41 \pm 2$ & $27 \pm 1$ & $...$ & $...$ \\
0221--055 & 24\,750 & 7.72 & 16.22 & 0.50 &  0.94 & 7.16 & $61 \pm 3$ & $42 \pm 2$ & $...$ & $...$ \\
0222--265 & 23\,200 & 7.91 & 15.68 & 0.58 &  1.74 & 7.43 & $76 \pm 4$ & $51 \pm 2$ & $...$ & $...$ \\
0227+050 & 19\,340 & 7.76 & 12.80 & 0.50 &  0.97 & 7.73 & $...$ & $775 \pm 38$ & $...$ & $255 \pm 12$ \\
0229+270 & 24\,160 & 7.90 & 15.52 & 0.58 &  1.71 & 7.31 & $79 \pm 4$ & $51 \pm 2$ & $...$ & $...$ \\
0231--054 & 17\,310 & 8.45 & 14.31 & 0.90 &  4.60 & 8.42 & $...$ & $236 \pm 11$ & $...$ & $71 \pm 3$ \\
0242--174 & 20\,660 & 7.85 & 15.38 & 0.55 &  1.41 & 7.66 & $96 \pm 4$ & $62 \pm 3$ & $...$ & $...$ \\
0300--232 & 22\,370 & 8.39 & 15.68 & 0.86 &  4.32 & 8.04 & $72 \pm 3$ & $44 \pm 2$ & $...$ & $...$ \\
0307+149 & 21\,410 & 7.91 & 15.38 & 0.58 &  1.69 & 7.62 & $112 \pm 5$ & $71 \pm 3$ & $...$ & $...$ \\
0308+188 & 18\,450 & 7.72 & 14.19 & 0.48 &  0.79 & 7.80 & $319 \pm 16$ & $204 \pm 10$ & $...$ & $...$ \\
0308--230 & 23\,570 & 8.54 & 15.08 & 0.96 &  5.18 & 8.10 & $108 \pm 5$ & $69 \pm 3$ & $...$ & $...$ \\
0331+226 & 21\,450 & 7.78 & 15.28 & 0.52 &  1.12 & 7.52 & $110 \pm 5$ & $70 \pm 3$ & $...$ & $...$ \\
0341+021 & 22\,150 & 7.27 & 15.41 & 0.33 &  ... & 7.27 & $101 \pm 5$ & $60 \pm 3$ & $...$ & $...$ \\
0349--256 & 20\,970 & 7.91 & 15.67 & 0.58 &  1.67 & 7.67 & $71 \pm 3$ & $47 \pm 2$ & $...$ & $...$ \\
0352+018 & 22\,110 & 7.80 & 15.57 & 0.52 &  1.19 & 7.46 & $79 \pm 4$ & $50 \pm 2$ & $...$ & $...$ \\
0358--514 & 23\,380 & 7.93 & 15.72 & 0.59 &  1.82 & 7.43 & $69 \pm 3$ & $47 \pm 2$ & $...$ & $...$ \\
0403--414 & 22\,700 & 7.94 & 16.35 & 0.60 &  1.85 & 7.51 & $41 \pm 2$ & $26 \pm 1$ & $...$ & $...$ \\
0410+117 & 21\,070 & 7.84 & 13.91 & 0.54 &  1.37 & 7.60 & $...$ & $246 \pm 12$ & $...$ & $63 \pm 3$ \\
0414--406 & 20\,940 & 8.00 & 16.13 & 0.63 &  2.13 & 7.75 & $47 \pm 2$ & $29 \pm 1$ & $...$ & $...$ \\
0416--105 & 24\,850 & 7.92 & 15.37 & 0.59 &  1.80 & 7.26 & $87 \pm 4$ & $52 \pm 2$ & $...$ & $...$ \\
0418--103 & 23\,390 & 8.29 & 15.68 & 0.80 &  3.76 & 7.87 & $70 \pm 3$ & $44 \pm 2$ & $...$ & $...$ \\
0421+162 & 19\,620 & 8.03 & 14.29 & 0.64 &  2.23 & 7.89 & $275 \pm 13$ & $171 \pm 8$ & $115 \pm 6$ & $60 \pm 5$ \\
0431+126 & 21\,370 & 7.97 & 14.23 & 0.61 &  1.98 & 7.68 & $...$ & $177 \pm 8$ & $...$ & $34 \pm 7$ \\
0452--347 & 21\,210 & 7.84 & 16.13 & 0.54 &  1.36 & 7.59 & $57 \pm 3$ & $36 \pm 1$ & $...$ & $...$ \\
0455--532 & 24\,430 & 7.55 & ... & 0.43 &  ... & 7.22 & $30 \pm 1$ & $17 \pm 1$ & $...$ & $...$ \\
0507+045.1 & 20\,840 & 7.90 & 14.22 & 0.57 &  1.62 & 7.67 & $...$ & $206 \pm 10$ & $...$ & $136 \pm 12$ \\
0843+516 & 23\,870 & 7.90 & 16.04 & 0.58 &  1.70 & 7.35 & $136 \pm 6$ & $137 \pm 7$ & $103 \pm 5$ & $162 \pm 9$ \\
0854+404 & 22\,250 & 7.91 & 14.81 & 0.58 &  1.71 & 7.53 & $156 \pm 7$ & $97 \pm 4$ & $...$ & $...$ \\
0859--039 & 23\,730 & 7.79 & 13.19 & 0.53 &  1.21 & 7.28 & $703 \pm 35$ & $435 \pm 21$ & $...$ & $...$ \\
0920+363 & 24\,060 & 7.63 & 16.07 & 0.46 &  0.58 & 7.22 & $56 \pm 2$ & $35 \pm 1$ & $...$ & $...$ \\
0933+025 & 22\,360 & 8.04 & 15.93 & 0.65 &  2.37 & 7.66 & $3569 \pm 178$ & $2450 \pm 122$ & $...$ & $...$ \\
0938+550 & 18\,530 & 8.10 & 14.79 & 0.68 &  2.62 & 8.04 & $197 \pm 9$ & $123 \pm 6$ & $...$ & $...$ \\
0944+192 & 17\,440 & 7.88 & 14.51 & 0.55 &  1.47 & 8.00 & $246 \pm 12$ & $159 \pm 8$ & $...$ & $...$ \\
0947+325 & 22\,060 & 8.31 & 15.50 & 0.82 &  3.87 & 7.98 & $87 \pm 4$ & $57 \pm 2$ & $...$ & $...$ \\
0954+697 & 21\,420 & 7.91 & 15.96 & 0.58 &  1.69 & 7.62 & $63 \pm 3$ & $44 \pm 2$ & $...$ & $...$ \\
1003--023 & 20\,610 & 7.89 & 15.27 & 0.57 &  1.58 & 7.69 & $120 \pm 6$ & $73 \pm 3$ & $...$ & $...$ \\
1005+642 & 19\,660 & 7.93 & 13.69 & 0.58 &  1.75 & 7.82 & $460 \pm 23$ & $291 \pm 14$ & $...$ & $...$ \\
1012--008 & 23\,200 & 8.07 & 15.59 & 0.67 &  2.57 & 7.62 & $79 \pm 4$ & $49 \pm 2$ & $...$ & $...$ \\
1013+256 & 21\,990 & 8.00 & 16.32 & 0.63 &  2.15 & 7.65 & $41 \pm 2$ & $25 \pm 1$ & $...$ & $...$ \\
1015+161 & 19\,950 & 7.92 & 15.61 & 0.58 &  1.73 & 7.78 & $196 \pm 9$ & $166 \pm 8$ & $146 \pm 7$ & $128 \pm 7$ \\
1017+125 & 21\,390 & 7.88 & 15.67 & 0.56 &  1.53 & 7.60 & $73 \pm 3$ & $48 \pm 2$ & $27 \pm 3$ & $18 \pm 6$ \\
1018+410 & 22\,390 & 8.04 & 16.37 & 0.65 &  2.37 & 7.66 & $85 \pm 4$ & $74 \pm 3$ & $...$ & $...$ \\
1020--207 & 19\,920 & 7.93 & 15.04 & 0.58 &  1.74 & 7.79 & $130 \pm 6$ & $83 \pm 4$ & $...$ & $...$ \\
1034+492 & 20\,650 & 8.17 & 15.43 & 0.73 &  3.05 & 7.94 & $100 \pm 5$ & $58 \pm 3$ & $48 \pm 3$ & $28 \pm 5$ \\
1038+633 & 24\,450 & 8.38 & 15.15 & 0.86 &  4.31 & 7.90 & $108 \pm 5$ & $67 \pm 3$ & $57 \pm 3$ & $17 \pm 4$ \\
1049+103 & 20\,550 & 7.91 & 15.81 & 0.58 &  1.67 & 7.71 & $3743 \pm 187$ & $2579 \pm 129$ & $...$ & $...$ \\
1049--158 & 20\,040 & 8.28 & 14.36 & 0.79 &  3.65 & 8.08 & $256 \pm 12$ & $155 \pm 7$ & $...$ & $...$ \\
1052+273 & 23\,100 & 8.37 & 14.12 & 0.86 &  4.23 & 7.98 & $304 \pm 15$ & $193 \pm 9$ & $129 \pm 7$ & $59 \pm 7$ \\
1058--129 & 24\,310 & 8.71 & 14.91 & 1.06 &  6.10 & 8.20 & $128 \pm 6$ & $81 \pm 4$ & $41 \pm 6$ & $29 \pm 8$ \\
1102+748 & 19\,710 & 8.36 & 15.05 & 0.84 &  4.13 & 8.18 & $136 \pm 6$ & $89 \pm 4$ & $...$ & $...$ \\
1104+602 & 17\,920 & 8.02 & 13.74 & 0.63 &  2.17 & 8.04 & $463 \pm 23$ & $293 \pm 14$ & $...$ & $...$ \\
1115+166 & 22\,090 & 8.12 & 15.05 & 0.70 &  2.80 & 7.77 & $133 \pm 6$ & $85 \pm 4$ & $69 \pm 4$ & $43 \pm 6$ \\
1122--324 & 21\,670 & 7.86 & 15.82 & 0.55 &  1.44 & 7.55 & $62 \pm 3$ & $40 \pm 2$ & $...$ & $...$ \\
1129+155 & 17\,740 & 8.03 & 14.09 & 0.64 &  2.22 & 8.06 & $376 \pm 18$ & $234 \pm 11$ & $159 \pm 8$ & $78 \pm 7$ \\
1133+293 & 23\,030 & 7.84 & 14.88 & 0.55 &  1.41 & 7.39 & $146 \pm 7$ & $90 \pm 4$ & $63 \pm 6$ & $43 \pm 8$ \\
1134+300 & 21\,280 & 8.55 & 12.45 & 0.96 &  5.23 & 8.26 & $1382 \pm 69$ & $889 \pm 44$ & $...$ & $...$ \\
1136+139 & 23\,920 & 7.83 & ... & 0.54 &  1.37 & 7.28 & $22 \pm 2$ & $15 \pm 1$ & $...$ & $...$ \\
1201--001 & 19\,770 & 8.26 & 15.16 & 0.78 &  3.55 & 8.08 & $124 \pm 6$ & $80 \pm 4$ & $48 \pm 3$ & $17 \pm 3$ \\
1204--322 & 21\,260 & 8.00 & 15.62 & 0.62 &  2.12 & 7.72 & $80 \pm 4$ & $52 \pm 2$ & $...$ & $...$ \\
1229--013 & 19\,430 & 7.47 & 14.46 & 0.38 &  ... & 7.54 & $239 \pm 12$ & $152 \pm 7$ & $90 \pm 5$ & $64 \pm 7$ \\
1230--308 & 22\,760 & 8.28 & 15.73 & 0.80 &  3.71 & 7.90 & $69 \pm 3$ & $45 \pm 2$ & $...$ & $...$ \\
1233--164 & 24\,890 & 8.21 & 15.10 & 0.76 &  3.33 & 7.66 & $114 \pm 5$ & $72 \pm 3$ & $...$ & $...$ \\
1243+015 & 21\,640 & 7.82 & 16.46 & 0.53 &  1.27 & 7.52 & $33 \pm 1$ & $23 \pm 1$ & $...$ & $...$ \\
1249+182 & 19\,910 & 7.73 & 15.24 & 0.49 &  0.85 & 7.65 & $95 \pm 4$ & $60 \pm 3$ & $...$ & $...$ \\
1257+048 & 21\,760 & 7.95 & 14.94 & 0.60 &  1.89 & 7.62 & $150 \pm 7$ & $97 \pm 4$ & $...$ & $...$ \\
1310--305 & 20\,350 & 7.82 & 14.48 & 0.53 &  1.25 & 7.66 & $223 \pm 11$ & $143 \pm 7$ & $...$ & $...$ \\
1323--514 & 19\,360 & 7.76 & 14.39 & 0.50 &  0.98 & 7.73 & $275 \pm 13$ & $175 \pm 8$ & $...$ & $...$ \\
1325+279 & 21\,270 & 8.04 & 15.80 & 0.65 &  2.35 & 7.76 & $63 \pm 7$ & $42 \pm 4$ & $...$ & $...$ \\
1325--089 & 17\,020 & 7.81 & ... & 0.52 &  1.12 & 8.00 & $165 \pm 8$ & $107 \pm 5$ & $...$ & $...$ \\
1330+473 & 22\,460 & 7.95 & 15.23 & 0.60 &  1.91 & 7.55 & $106 \pm 5$ & $70 \pm 3$ & $...$ & $...$ \\
1334--160 & 18\,650 & 8.32 & ... & 0.82 &  3.86 & 8.20 & $116 \pm 6$ & $98 \pm 5$ & $...$ & $...$ \\
1335+369 & 20\,510 & 7.78 & ... & 0.51 &  1.08 & 7.62 & $2056 \pm 102$ & $1326 \pm 66$ & $...$ & $...$ \\
1337+705 & 20\,460 & 7.90 & 12.60 & 0.57 &  1.62 & 7.71 & $...$ & $709 \pm 35$ & $...$ & $254 \pm 13$ \\
1338+081 & 24\,440 & 7.65 & 16.44 & 0.47 &  0.66 & 7.20 & $30 \pm 1$ & $19 \pm 1$ & $...$ & $...$ \\
1353+409 & 23\,530 & 7.59 & 15.49 & 0.44 &  ... & 7.26 & $76 \pm 3$ & $50 \pm 2$ & $32 \pm 4$ & $26 \pm 5$ \\
1408+323 & 18\,150 & 7.95 & 13.97 & 0.59 &  1.81 & 7.97 & $...$ & $251 \pm 12$ & $...$ & $92 \pm 7$ \\
1433+538 & 22\,410 & 7.80 & 16.14 & 0.53 &  1.21 & 7.43 & $906 \pm 45$ & $632 \pm 31$ & $...$ & $...$ \\
1449+168 & 22\,350 & 7.79 & 15.39 & 0.52 &  1.15 & 7.42 & $91 \pm 4$ & $58 \pm 2$ & $...$ & $...$ \\
1451+006 & 25\,480 & 7.89 & 15.27 & 0.58 &  1.69 & 7.16 & $98 \pm 5$ & $63 \pm 3$ & $...$ & $...$ \\
1452--042 & 23\,530 & 8.19 & 16.29 & 0.74 &  3.21 & 7.74 & $51 \pm 2$ & $36 \pm 2$ & $...$ & $...$ \\
1457--086 & 21\,450 & 7.97 & 15.76 & 0.61 &  1.99 & 7.68 & $113 \pm 5$ & $73 \pm 3$ & $48 \pm 3$ & $...$ \\
1459+347 & 21\,520 & 8.48 & 15.79 & 0.92 &  4.84 & 8.18 & $65 \pm 3$ & $41 \pm 2$ & $...$ & $...$ \\
1507+220 & 19\,870 & 7.75 & 14.95 & 0.49 &  0.91 & 7.66 & $150 \pm 7$ & $96 \pm 4$ & $...$ & $...$ \\
1524--749 & 23\,090 & 7.74 & 15.99 & 0.50 &  0.99 & 7.32 & $64 \pm 3$ & $39 \pm 2$ & $...$ & $...$ \\
1525+257 & 22\,290 & 8.28 & 15.73 & 0.80 &  3.70 & 7.93 & $69 \pm 3$ & $44 \pm 2$ & $...$ & $...$ \\
1527+090 & 21\,200 & 7.85 & 14.30 & 0.55 &  1.39 & 7.59 & $259 \pm 13$ & $164 \pm 8$ & $...$ & $...$ \\
1531--022 & 18\,620 & 8.41 & 13.97 & 0.87 &  4.41 & 8.30 & $...$ & $243 \pm 12$ & $...$ & $103 \pm 8$ \\
1533--057 & 20\,000 & 8.50 & 15.38 & 0.93 &  4.94 & 8.29 & $94 \pm 4$ & $57 \pm 2$ & $...$ & $...$ \\
1535+293 & 24\,470 & 7.92 & ... & 0.59 &  1.81 & 7.30 & $50 \pm 2$ & $30 \pm 1$ & $...$ & $...$ \\
1547+057 & 24\,360 & 8.36 & 15.94 & 0.85 &  4.16 & 7.88 & $56 \pm 2$ & $36 \pm 1$ & $...$ & $...$ \\
1548+149 & 21\,450 & 7.86 & 15.16 & 0.55 &  1.45 & 7.57 & $122 \pm 6$ & $77 \pm 3$ & $...$ & $...$ \\
1614+136 & 22\,020 & 7.21 & 15.23 & 0.32 &  ... & 7.23 & $113 \pm 5$ & $67 \pm 3$ & $54 \pm 5$ & $15 \pm 5$ \\
1620--391 & 24\,680 & 7.93 & 11.00 & 0.60 &  1.85 & 7.29 & $4988 \pm 249$ & $3113 \pm 155$ & $1936 \pm 96$ & $1073 \pm 53$ \\
1633+676 & 23\,660 & 7.97 & 16.25 & 0.62 &  2.04 & 7.45 & $34 \pm 1$ & $21 \pm 1$ & $...$ & $...$ \\
1647+375 & 21\,980 & 7.89 & 14.91 & 0.57 &  1.61 & 7.55 & $137 \pm 6$ & $86 \pm 4$ & $...$ & $...$ \\
1713+332 & 22\,120 & 7.43 & 14.39 & 0.38 &  ... & 7.32 & $...$ & $184 \pm 9$ & $...$ & $57 \pm 6$ \\
1755+194 & 24\,440 & 7.80 & 15.99 & 0.53 &  1.28 & 7.21 & $54 \pm 3$ & $33 \pm 1$ & $...$ & $...$ \\
1914--598 & 19\,760 & 7.84 & ... & 0.54 &  1.31 & 7.74 & $288 \pm 14$ & $196 \pm 10$ & $...$ & $...$ \\
1929+012 & 20\,890 & 7.91 & 14.20 & 0.58 &  1.68 & 7.68 & $837 \pm 41$ & $781 \pm 39$ & $...$ & $...$ \\
1943+163 & 19\,760 & 7.79 & 13.95 & 0.51 &  1.11 & 7.71 & $...$ & $230 \pm 11$ & $...$ & $100 \pm 7$ \\
1953--715 & 19\,270 & 7.87 & 15.06 & 0.55 &  1.47 & 7.81 & $132 \pm 6$ & $84 \pm 4$ & $...$ & $...$ \\
2021--128 & 20\,750 & 7.82 & 15.20 & 0.53 &  1.26 & 7.62 & $112 \pm 5$ & $74 \pm 3$ & $...$ & $...$ \\
2032+188 & 18\,200 & 7.36 & 15.32 & 0.34 &  ... & 7.58 & $125 \pm 6$ & $80 \pm 4$ & $...$ & $...$ \\
2039--202 & 19\,740 & 7.79 & 12.33 & 0.51 &  1.08 & 7.70 & $...$ & $1035 \pm 51$ & $...$ & $371 \pm 20$ \\
2046--220 & 23\,410 & 7.83 & 15.37 & 0.54 &  1.36 & 7.33 & $88 \pm 4$ & $55 \pm 2$ & $...$ & $...$ \\
2058+181 & 17\,350 & 7.75 & 15.20 & 0.49 &  0.88 & 7.94 & $149 \pm 7$ & $97 \pm 5$ & $...$ & $...$ \\
2134+218 & 18\,000 & 7.86 & 14.48 & 0.55 &  1.39 & 7.93 & $...$ & $158 \pm 8$ & $...$ & $61 \pm 6$ \\
2149+021 & 17\,930 & 7.86 & 12.76 & 0.54 &  1.38 & 7.94 & $...$ & $776 \pm 38$ & $...$ & $272 \pm 15$ \\
2152--045 & 19\,840 & 7.38 & 15.69 & 0.35 &  ... & 7.44 & $83 \pm 4$ & $52 \pm 2$ & $...$ & $...$ \\
2200--136 & 24\,730 & 7.61 & 15.33 & 0.45 &  0.53 & 7.19 & $106 \pm 5$ & $66 \pm 3$ & $...$ & $...$ \\
2204+071 & 24\,450 & 7.95 & 15.78 & 0.61 &  1.95 & 7.34 & $68 \pm 3$ & $41 \pm 2$ & $...$ & $...$ \\
2210+233 & 23\,230 & 8.24 & 15.84 & 0.77 &  3.48 & 7.82 & $59 \pm 3$ & $34 \pm 1$ & $...$ & $...$ \\
2220+133 & 22\,580 & 8.30 & 15.61 & 0.81 &  3.82 & 7.93 & $75 \pm 3$ & $45 \pm 2$ & $...$ & $...$ \\
2220+217.1 & 18\,740 & 8.24 & 15.22 & 0.77 &  3.43 & 8.13 & $79 \pm 4$ & $45 \pm 2$ & $...$ & $...$ \\
2229+235 & 19\,300 & 7.90 & 16.01 & 0.57 &  1.60 & 7.83 & $59 \pm 3$ & $38 \pm 2$ & $23 \pm 2$ & $...$ \\
2231--267 & 21\,590 & 7.99 & 14.97 & 0.62 &  2.09 & 7.68 & $162 \pm 8$ & $108 \pm 5$ & $...$ & $...$ \\
2238--045 & 17\,540 & 8.18 & 16.90 & 0.73 &  3.04 & 8.17 & $22 \pm 1$ & $13 \pm 0$ & $...$ & $...$ \\
2244+210 & 24\,110 & 7.89 & 16.45 & 0.57 &  1.66 & 7.31 & $35 \pm 1$ & $21 \pm 1$ & $...$ & $...$ \\
2257+162 & 24\,580 & 7.49 & 15.97 & 0.41 &  ... & 7.24 & $417 \pm 20$ & $287 \pm 14$ & $195 \pm 10$ & $128 \pm 8$ \\
2306+124 & 20\,360 & 7.99 & 15.08 & 0.62 &  2.08 & 7.80 & $124 \pm 6$ & $76 \pm 3$ & $...$ & $...$ \\
2322--181 & 21\,680 & 7.90 & 15.28 & 0.57 &  1.65 & 7.59 & $99 \pm 5$ & $61 \pm 3$ & $...$ & $...$ \\
2328+107 & 21\,000 & 7.78 & 15.76 & 0.51 &  1.08 & 7.56 & $115 \pm 5$ & $77 \pm 3$ & $...$ & $...$ \\
2359--324 & 22\,480 & 7.74 & 16.26 & 0.50 &  0.97 & 7.38 & $43 \pm 2$ & $27 \pm 1$ & $...$ & $...$ \\
\hline
\label{ref:allstars}
\end{longtable}
\twocolumn

\label{lastpage}

\end{document}